\documentclass[aps,preprint,amsmath,amssymb,groupedaddress,superscriptaddress]{revtex4-1}

\pdfoutput=1

\usepackage{amsmath}
\usepackage{amssymb}

\usepackage{graphicx}
\usepackage{subfigure}
\usepackage{longtable}

\usepackage{dcolumn}
\usepackage{bm}
\usepackage{array}
\usepackage{color}

\usepackage{longtable}

\graphicspath{{../1figures/}}

\begin{document}

\title{Two faces of greedy leaf removal procedure on graphs}

\author{Jin-Hua Zhao}
\email{E-mail: zhaojh190@gmail.com}

\affiliation{
CAS Key Laboratory of Theoretical Physics,
Institute of Theoretical Physics,
Chinese Academy of Sciences,
Beijing 100190,
China}

\affiliation{
Institute of Quantum Matter and
School of Physics and Telecommunication Engineering,
South China Normal University,
Guangzhou 510006, China}

\author{Hai-Jun Zhou}

\affiliation{
CAS Key Laboratory of Theoretical Physics,
Institute of Theoretical Physics,
Chinese Academy of Sciences,
Beijing 100190,
China}

\affiliation{
School of Physical Sciences,
University of Chinese Academy of Sciences,
Beijing 100049,
China}

\date{\today}


\begin{abstract}
The greedy leaf removal (GLR) procedure on a graph
is an iterative removal of any vertex with degree one (leaf)
along with its nearest neighbor (root).
Its result has two faces: a residual subgraph as a core, and a set of removed roots.
While the emergence of cores
on uncorrelated random graphs was solved analytically,
a theory for roots is ignored
except in the case of Erd\"{o}s-R\'{e}nyi random graphs.
Here we analytically study roots on random graphs.
We further show that,
with a simple geometrical interpretation and a concise mean-field theory of the GLR procedure,
we reproduce the zero-temperature replica symmetric estimation
of relative sizes of both minimal vertex covers and maximum matchings on random graphs
with or without cores.
\end{abstract}

\maketitle

\tableofcontents

\newpage

\section{Introduction}

As a quantitative approach
to model the topology of interconnection and interaction
among the constituents of complex systems in various nature,
the paradigm of graphs and networks
\cite{
Bollobas-2002,
Albert.Barabasi-RMP-2002,
Newman-SIAM-2003,
Boccaleti.Latora.Moreno.Chavez.Hwang-PhysRep-2006,
Dorogovtsev.Goltsev-RMP-2008,
Newman-2018}
is a simple choice.
Its two major topics are
the percolation phenomena and the combinatorial optimization.
A percolation problem on a graph
\cite{Stauffer.Aharony-1994},
which tackles intrinsically geometrical transitions,
focuses on the subgraph size
after a local procedure of vertex and edge removal
(equivalently a vertex and edge addition to an empty graph).
Its analysis provides a structural perspective to the stability
of interconnected systems upon internal fluctuation and external perturbation,
further sheds light on the management of critical transitions
in high-dimensional dynamical systems
\cite{Scheffer-2009}.
A well-known percolation problem
is the emergence of the giant connected component (GCC)
\cite{
Molloy.Reed-RandStructAlgo-1995,
Albert.Jeong.Barabasi-Nature-2000,
Cohen.etal-PRL-2000,
Callaway.etal-PRL-2000}.
A combinatorial optimization problem on a graph
\cite{
Papadimitriou.Steiglitz-1998,
Pardalos.Du.Graham-2013-2e}
usually concerns finding
an optimal set of vertices or edges under a structural constraint.
The difficulty in finding solutions
leads to the study of computational complexity classes
\cite{Garey.Johnson-1979},
whose typical examples are the polynomial
and the non-deterministic polynomial-time-hard (NP-hard) problems.
From an algorithmic sense,
a polynomial problem has a proper solver with a computation time $t$
as a polynomial function of the underlying graph size $N$
($t \sim N^{c}$ with a finite constant $c$),
while finding a solution of a NP-hard problem in the worst case
needs an exponential computation time in order of the underlying graph size $N$
($t \sim e ^{dN}$ with a finite constant $d$).

The focus of the paper is at the intersection between
the percolation and the combinatorial optimization problems.
Specifically,
we consider the implication of the greedy leaf removal (GLR) procedure
in the minimal vertex cover (MVC) and the maximum matching (MM) problems on undirected graphs.
The MVC problem,
which is a NP-hard problem,
concerns finding vertex covers of a graph
(sets of vertices to which all the edges of the graph are adjacent)
of the minimal cardinality.
A list of results is derived with exact methods
\cite{
Gazmuri-Network-1984,
Harant-DiscreteMath-1998,
Frieze-DiscreteMath-1990},
numerical approximations
\cite{Takabe.Hukushima-JStatMech-2016}, and
statistical physical methods
(replica trick and cavity method)
\cite{
Mezard.Parisi.Virasoro-1987,
Nishimori-2001,
Mezard.Montanari-2009,
Weigt.Hartmann-PRL-2000,
Weigt.Hartmann-PRL-2001,
Weigt.Hartmann-PRE-2001,
Zhou-EPJB-2003,
Zhou-PRL-2005,
Weigt.Zhou-PRE-2006,
Zhou.Zhou-PRE-2009,
Zhang.Zeng.Zhou-PRE-2009}.
Reviews with a statistical physical background can be found in
\cite{
Hartmann.Weigt-2005,
Hartmann.Weigt-JPhyA-2003,
Zhao.Zhou-CPB-2014}.
The MM problem
\cite{Lovasz.Plummer-1986},
which is a polynomial problem,
concerns finding matchings on a graph
(sets of edges sharing no common vertex)
of the maximal size.
Treatments of the problem with statistical physical methods
\cite{
Zhou.OuYang-arxiv-2003,
Zdeborova.Mezard-JStatMech-2006}
can be found therein.
The GLR procedure on a graph
\cite{
Karp.Sipser-IEEFoCS-1981,
Aronson.Frieze.Pittel-RandStrucAlgo-1998}
iteratively removes any vertex with only one nearest neighbor (leaf)
along with its sole nearest neighbor (root).
It results in core percolation,
whose mean-field theory on uncorrelated random graphs is developed in
\cite{
Bauer.Golinelli-EPJB-2001,
Liu.Csoka.Zhou.Posfai-PRL-2012}.
The GLR procedure is adopted as a local step to approximate MVCs and MMs
\cite{
Karp.Sipser-IEEFoCS-1981,
Aronson.Frieze.Pittel-RandStrucAlgo-1998,
Bauer.Golinelli-EPJB-2001},
and the core percolation corresponds to a fundamental transition
in the organization of solution spaces of both optimization problems
\cite{
Weigt.Hartmann-PRL-2000,
Weigt.Hartmann-PRL-2001,
Weigt.Hartmann-PRE-2001,
Zhou-EPJB-2003,
Zhou-PRL-2005,
Weigt.Zhou-PRE-2006,
Zhou.Zhou-PRE-2009,
Zhang.Zeng.Zhou-PRE-2009,
Zhou.OuYang-arxiv-2003,
Zdeborova.Mezard-JStatMech-2006}.
The GLR procedure variants for combinatorial optimizations
with concurrent percolation phenomena and solution space transitions
also show in the MM problem of the network controllability
\cite{
Liu.Slotine.Barabasi-Nature-2011,
Jia.etal-NatComm-2013},
the $p$-spin model and the XOR-SAT problem
\cite{
Mezard.RicciTersenghi.Zecchina-JStatPhys-2003,
Cocco.etal-PRL-2003},
the Boolean networks
\cite{
Correale.etal-PRL-2006,
Correale.etal-JStatMech-2006},
the maximum set packing problem
\cite{Lucibello.RicciTersenghi-IntJStatMech-2014},
and the minimum dominating set problem
\cite{
Haynes.Hedetniemi.Slater-1998,
Zhao.Zhou-JStatPhys-2015,
Zhao.Zhou-LNCS-2015,
Habibulla-JStatMech-2017}.

The GLR procedure leaves a core and a set of roots.
A scenario with two faces also shows in
the articulation points
\cite{
Tarjan.Vishkin-SIAMJComput-1985,
Liu.etal-NatCommun-2017}
and the $k$-core pruning process
\cite{
Chalupa.Leath.Reich-JPhysC-1979,
Dorogovtsev.etal-PRL-2006,
Baxter.etal-PRX-2015}.
It is established that
\cite{
Karp.Sipser-IEEFoCS-1981,
Bauer.Golinelli-EPJB-2001}:
(1) on a graph without core,
from roots we can reconstruct a MVC and a MM;
correspondingly, the root size is just the MVC or MM size;
(2) on a general graph,
with its roots and core,
we can estimate an average MM size
based on the perfect matching of cores
(each vertex in a core is adjacent to a matched edge).
Yet unlike the core,
the roots lack a general analytical treatment
except on the Erd\"{o}s-R\'{e}nyi random graphs
\cite{
Karp.Sipser-IEEFoCS-1981,
Bauer.Golinelli-EPJB-2001},
and the geometrical implication of roots  and cores on estimating MVC and MM sizes is not fully explored.
Here we sum our main contribution in this paper:
(1) we complete the missing piece of an analytical theory of root sizes on uncorrelated random graphs;
(2) based on the geometrical interpretation and the analytical theory of roots and cores,
we develop simple frameworks to estimate MVC and MM fractions on random graphs,
which retrieve their zero-temperature replica symmetric results from previous statistical physical literature.

The layout of the paper
is as follows.
In section \ref{sec:model},
we explain the GLR procedure, the MVC problem, and the MM problem.
In section \ref{sec:theory},
we lay down a mean-field theory for the two faces of the GLR procedure
on uncorrelated random graphs,
and its implication in estimating MVC and MM sizes.
In section \ref{sec:results},
we test our framework on some random graph models and real networks.
In section \ref{sec:conclusion},
we discuss and conclude the paper.

\section{Model}
\label{sec:model}

\begin{figure}
\begin{center}
 \includegraphics[width = 0.70 \linewidth]{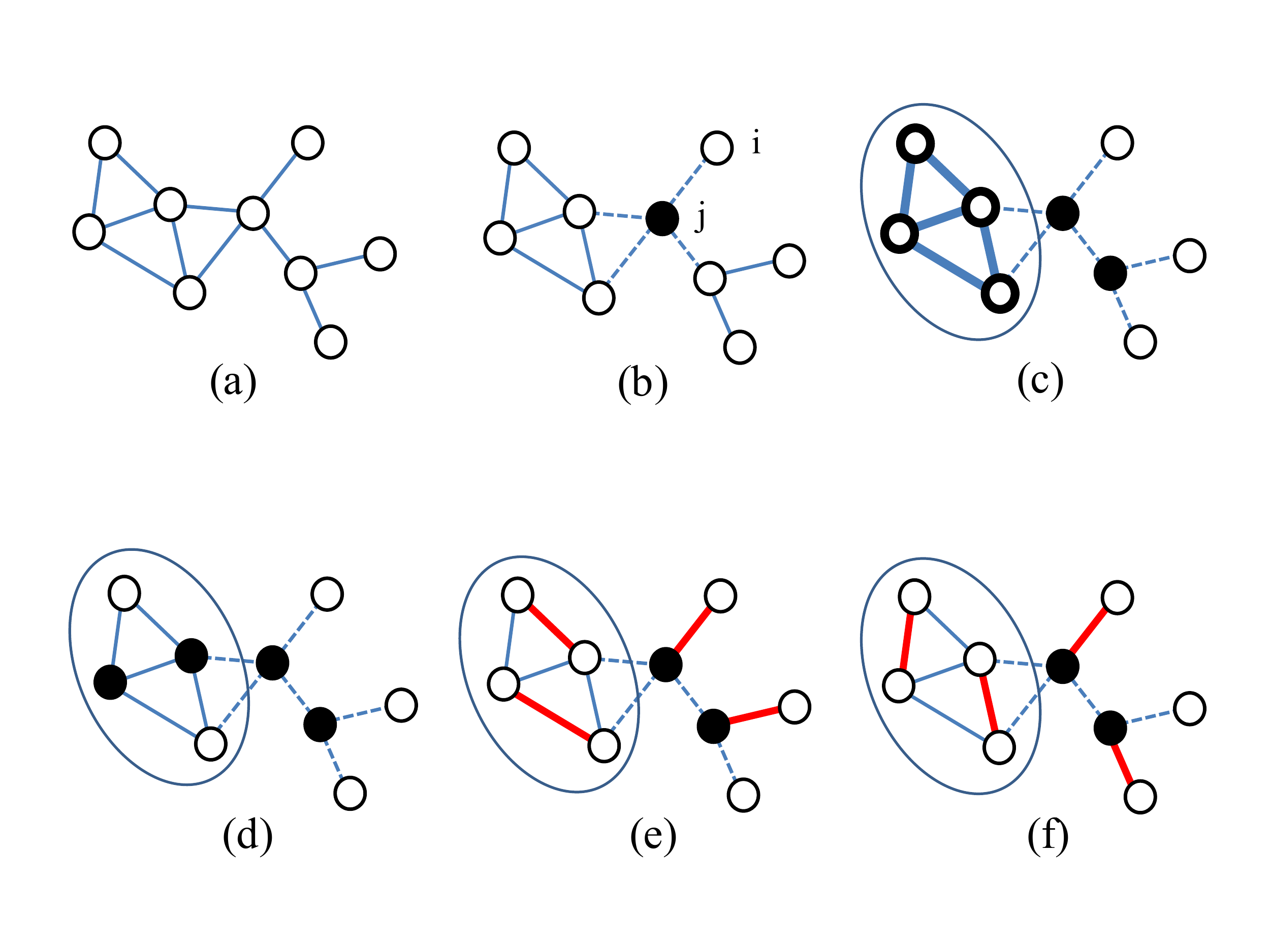}
\end{center}
\caption{
 \label{fig:glr_model}
The GLR procedure, the MVCs, and the MMs on a small graph.
(a) shows a graph with $9$ vertices and $11$ edges.
(b) shows a single GLR step,
in which the leaf $i$ and its neighboring root $j$ (in a filled circle)
are removed along with their adjacent edges
(in dashed lines).
(c) shows the result of the GLR procedure,
a set of $2$ roots (in filled circles) and a core consisting of $4$ vertices and $5$ edges
(enclosed in a large circle).
(d) shows a MVC as a set of $4$ vertices (in filled circles).
(e) and (f) show two MMs both as a set of $4$ edges (in thick solid lines).}
\end{figure}
%

First we explain some notions for graphs.
An undirected graph $G = \{V, E\}$
consists of a vertex set $V$ and an edge set $E \subset V \times V$.
For an edge $(i, j)$ between the end-vertices $i$ and $j$,
$i$ is a nearest neighbor of $j$, and vice versa.
The degree $k_{i}$ of a vertex $i$
is the size of its nearest neighbors or $|\partial i|$.
The degree distribution $P(k)$
is the probability of finding a randomly chosen vertex with a degree $k$.
The mean degree or connectivity $c$ is the average degree of a vertex,
or $c \equiv 2 |E| / |V| = \sum _{k = 0}^{\infty} k P(k)$.
The excess degree distribution $Q(k)$
is the probability of arriving a vertex with a degree $k$
following a randomly chosen edge,
equivalently $Q(k) = k P(k) / c$.
We also define cavity graphs for a graph.
The cavity graph $G \backslash i$
is the subgraph of $G$ with the vertex $i$ and all its adjacent edges removed,
equivalently $\{V', E'\}$
with $V' = V \backslash i$ and $E' = E \backslash \{(k, i)\}$ with $k \in \partial i$.
The cavity graph $G \backslash (i, j)$
is the subgraph of $G$ with the edge $(i, j)$ removed,
equivalently $\{V, E^{'}\}$ with $E^{'} = E \backslash (i, j)$.

The GLR procedure consists of single iterative GLR steps
in which a leaf and its neighboring root is removed along with all their adjacent edges,
and finally leaves a subgraph as the core.
Figures \ref{fig:glr_model} (a) - (c) show an example.
We should mention that
\cite{Bauer.Golinelli-EPJB-2001}:
the core of a graph is well defined,
or the vertices and the edges in a core
are independent of the GLR pruning process;
the set of roots depends on the trajectory of the pruning process,
yet on large graphs its size converges on average.

A vertex cover (VC) of $G$
is a set of vertices $\Gamma \subset V$
to which each edge in $E$ is adjacent.
Taking the GLR procedure as a local method,
in each step the root (such as the vertex $j$ in figure \ref{fig:glr_model} (b))
is selected into a VC
and all its adjacent edges are removed as satisfied constraints.
If there is no core after the GLR procedure,
the set of roots is simply a MVC;
otherwise, local algorithms or statistical physical methods
can be further applied on the core to approximate a MVC.
Figure \ref{fig:glr_model} (d) shows an example.

A matching of $G$
is a set of edges $\Delta \subset E$ without shared vertices.
The GLR procedure serves as a component of the well-known Karp-Sipser algorithm
\cite{Karp.Sipser-IEEFoCS-1981},
a randomized algorithm which finds MMs with a high probability on large graphs.
The Karp-Sipser algorithm on a graph goes as:
(1) the GLR procedure is applied on the current graph,
while in each step the edge bridging a leaf and its neighboring root is selected into a matching;
(2) an edge in the core is randomly selected into the matching,
and any edge adjacent to its end-vertices is removed;
(3) the above two steps are iteratively carried out until there is no edge left,
and the resulted matching is an approximate MM.
Figure \ref{fig:glr_model} (e) and (f) show an example.
If there is no core left on a graph,
the MM size is simply the root size;
otherwise,
we can adopt the perfect matching of cores
\cite{Karp.Sipser-IEEFoCS-1981},
which assumes that any vertex in a core is adjacent to a certain edge in the MM of the core.
Taken together,
the MM size on a large graph can be estimated conveniently as the root size plus half the core vertex size.

\section{Theory}
\label{sec:theory}

As we can see,
the GLR procedure applies on any undirected graph instance.
Yet on uncorrelated random graphs,
we can analytically calculate the size of roots
by extending a cavity method for cores from the GLR procedure in
\cite{Liu.Csoka.Zhou.Posfai-PRL-2012}.
By further exploring the geometrical meaning of the roots and cores,
we can develop analytical frameworks for both the MVC and MM sizes
on random graphs.

\subsection{GLR}

On an uncorrelated random graph $G = \{V, E\}$
with a vertex set $V$ and an edge set $E$,
we define two cavity probabilities tailored for the GLR procedure.
Following a randomly chosen edge $(i, j) \in E$ from the vertex $i$ to the vertex $j$,
we define $\alpha$ ($\beta$)
as the probability that $j$ becomes a leaf (a root) in the GLR procedure on the cavity graph $G \backslash (i, j)$.
On sparse random graphs,
the local tree-structure approximation for the belief propagation algorithms
\cite{
Mezard.Montanari-2009,
Bethe-ProcRSocLondA-1935,
Kschischang.Frey.Loeliger-IEEETransInfTheor-2001,
Yedidia.Freeman.Weiss-IEEETransInfTheor-2005}
assumes that the states of the nearest neighbors of any vertex, say $i \in V$,
are independent of each other in the pruning process
on the cavity graph $G \backslash i$.
Based on this approximation,
we have the self-consistent equations of $\alpha$ and $\beta$ as
\begin{eqnarray}
\label{eq:alpha}
\alpha
& = &
\sum _{k = 1}^{\infty}
Q(k) \beta ^{k - 1}, \\
\label{eq:beta}
\beta
& = &
1- \sum _{k = 1}^{\infty}
Q(k) (1 - \alpha)^{k - 1}.
\end{eqnarray}
%
We define $n$ as the fraction of vertices in a core,
$l$ the relative size of edges in a core normalized by the vertex size $|V|$,
and $w$ the fraction of roots.
In the framework of cavity method,
these relative sizes can be calculated as marginal probabilities
with the stable solutions of $\alpha$ and $\beta$.
We have
\begin{eqnarray}
\label{eq:n}
n
&&
= \sum _{k  = 2}^{\infty}
P(k) \sum _{s = 2}^{k}
 \left(\begin{array}{c} k \\ s \end{array}\right)
\beta ^{k -s}
(1 - \alpha - \beta)^{s}, \\
\label{eq:l}
l
&&
= \frac {1}{2} c (1 - \alpha - \beta)^2, \\
\label{eq:w}
w
& &
= 1 - \sum _{k = 0}^{\infty} P(k) (1 - \alpha) ^{k}
- \frac {1}{2} c \alpha ^{2}.
\end{eqnarray}
Equation (\ref{eq:n}) can be reformulated as
$n = \sum _{k = 0}^{\infty} P(k) [(1 - \alpha)^{k} - \beta ^{k}] - c \alpha (1 - \alpha - \beta)$.

Equations (\ref{eq:alpha}) and (\ref{eq:beta})
have been derived in
\cite{Zdeborova.Mezard-JStatMech-2006}
and equations (\ref{eq:alpha}) - (\ref{eq:l}) in
\cite{Liu.Csoka.Zhou.Posfai-PRL-2012},
yet equation (\ref{eq:w}) is our contribution here.
Here is a simple explanation for them.
For equations (\ref{eq:alpha}) and (\ref{eq:beta}),
we consider the case from a cavity graph $G \backslash j$
to another cavity graph $G \backslash (i, j)$
after some edge addition,
in which $(i, j) \in E$ is a randomly chosen edge.
If $j$ is a leaf on $G \backslash (i, j)$,
all its nearest neighbors except $i$ must be removed as roots on $G \backslash j$,
thus we have equation (\ref{eq:alpha}).
In a similar sense,
if the vertex $j$ is a root on $G \backslash (i, j)$,
there must be at least one nearest neighbors except $i$ turning into leaves on $G \backslash j$,
thus we have equation (\ref{eq:beta}).
We then consider the case from a cavity graph $G \backslash i$
to the original graph $G$
after some edge addition,
in which $i \in V$ is a randomly chosen vertex.
If a newly added vertex $i$ is in the core of $G$,
among all the nearest neighbors of $i$,
there must be no leaf, only roots,
and at least two vertices in the core of $G \backslash i$
to forbid the GLR procedure,
thus we have equation (\ref{eq:n}).
If a newly added vertex $i$ is a root of $G$,
among all the nearest neighbors of $i$ on $G \backslash i$,
there must be at least one leaves,
thus we have the first two terms on the right-hand side (RHS) of equation (\ref{eq:w}).
Yet there is a recounting case.
We consider the case from a cavity graph $G \backslash (i, j)$ to $G$,
in which $(i, j) \in E$ is a randomly chosen edge.
If $i$ and $j$ are both leaves on $G \backslash (i, j)$,
they could be both counted as roots on $G$.
The coefficient $c / 2 (= |E| / |V|)$ is the relative density of edges to vertices.
Thus we have the third term on the RHS of equation (\ref{eq:w}).
For equation (\ref{eq:l}),
we also consider the case from $G \backslash (i, j)$ to $G$.
If a newly added edge $(i, j)$ is in the core of $G$,
both $i$ and $j$ must be in the core of $G \backslash (i, j)$.
The coefficient $c / 2$ follows the same logic in equation (\ref{eq:w}).

In appendix A,
other than the cumulative approach in this section,
we provide a discrete approach to the mean-field theory of the GLR procedure.
This approach is based on the stepwise interpretation of the pruning process:
at each time-step with an index $t \ge 0$,
all the leaves are determined,
their neighboring roots selected,
and all their adjacent edges removed;
the discrete steps are carried out iteratively until there is no leaf left at large $t$.
The discrete viewpoint of a local process on graphs
is adopted in various percolation and optimization problems
\cite{
Bauer.Golinelli-EPJB-2001,
Zhao.Zhou-JStatPhys-2015,
Zhao.Zhou-LNCS-2015,
Liu.etal-NatCommun-2017,
Baxter.etal-PRX-2015}.
With the logic in equations (\ref{eq:alpha}) and (\ref{eq:beta}),
following a randomly chosen edge $(i, j) \in E$ from the vertex $i$ to the vertex $j$,
we define $\alpha ^{(t)}$ ($\beta ^{(t)}$) with $t \ge 0$
as the cavity probabilities of $j$ to be a leaf (a root)
at the $t$-th time-step of the discretized GLR procedure on the cavity graph $G \backslash (i, j)$.
We can further calculate the relative sizes of the removed roots and the residual subgraph size
at each time-step.
In appendix A,
we will show that in the case of infinitely large time-steps ($t \rightarrow \infty$),
the discrete formulation
reduces to the cumulative one in this section.

\subsection{MVC}

For the MVC problem on a graph,
we denote $x$ as the relative size of a MVC
(the size of vertices in a MVC normalized by the graph vertex size).
It is an established result
\cite{
Karp.Sipser-IEEFoCS-1981,
Bauer.Golinelli-EPJB-2001} that:
if a graph has no core,
the roots constitute a MVC;
otherwise,
the size of the roots provides a lower bound for MVCs.
From an algorithmic perspective,
on any graph instance we have
\begin{eqnarray}
x =
\left \{ \begin{array}{rcl}
w,
& \mbox{for} & n = 0; \\
w + x_{core},
& \mbox{for} & n > 0,
\end{array}\right.
\end{eqnarray}
while $x_{core}$ is the MVC size of the core,
which can be approximated efficiently with message-passing algorithms
\cite{
Weigt.Zhou-PRE-2006,
Zhao.Zhou-CPB-2014}.

Besides the above obvious result,
we can go deeper into the region of graphs with cores.
An analysis of the fixed points of $\alpha$ and $\beta$
from equations (\ref{eq:alpha}) and (\ref{eq:beta})
on infinitely large graphs
shows that:
before the core percolation,
there is only one branch of fixed points as the stable solution,
denoted as $(\alpha _{\textrm{M}}, \beta _{\textrm{M}})$
with $1 - \alpha _{\textrm{M}} - \beta _{\textrm{M}} = 0$;
after the core percolation,
there are another two new branches of fixed points emerged
as the lower branch $(\alpha _{\textrm{L}}, \beta _{\textrm{L}})$
and the upper branch $(\alpha _{\textrm{U}}, \beta _{\textrm{U}})$
with $\alpha _{\textrm{L}} < \alpha _{\textrm{M}} < \alpha _{\textrm{U}}$ and
$1 - \alpha _{\textrm{L}} - \beta _{\textrm{L}} = - (1 - \alpha _{\textrm{U}} - \beta _{\textrm{U}}) > 0$,
among which the lower branch of $(\alpha _{\textrm{L}}, \beta_{\textrm{L}})$ acts as the stable one.
Yet, the middle branch with $1 - \alpha _{\textrm{M}} - \beta _{\textrm{M}} = 0$
is always a solution of equations (\ref{eq:alpha}) and (\ref{eq:beta}),
under which $n = 0$ and
$w$ gives an estimation of $x$ on graphs with or without cores.
We consider the solution branch with $1 - \alpha - \beta = 0$
(dropping the subscripts conveniently)
as the trivial core condition,
and the corresponding $w$ as the trivial core solution of the MVC problem.
Taken together,
from an analytical perspective,
on large random graphs we have
\begin{eqnarray}
\label{eq:mvc_rs}
x = w,\ \rm{s.t.} \ 1 - \alpha - \beta = 0. 
\end{eqnarray}
Equations (\ref{eq:alpha}), (\ref{eq:beta}), (\ref{eq:w}), and (\ref{eq:mvc_rs})
constitute our framework to estimate the MVC sizes on random graphs
with or without cores.
On random graphs without cores,
our framework gives the exact MVC sizes;
otherwise,
our framework is an underestimation of the MVC sizes.
Later we will see that our framework simply reproduces
the well-known zero-temperature replica symmetric estimation
of the MVC sizes on Erd\"{o}s-R\'{e}nyi random graphs in
\cite{Weigt.Hartmann-PRL-2000}.

We further estimate sizes of backbones in MVC solutions.
With the notation in
\cite{Weigt.Hartmann-PRL-2000},
we define the fraction of covered backbone $b_{+}$
as the relative size of vertices which are always in the MVC solutions,
the fraction of uncovered backbone $b_{-}$
the relative size of vertices which are absent from any MVC solution.
From a geometrical viewpoint,
the covered backbone
is simply the set of the roots
excluding those from the isolated edges
whose both end-vertices can be in a MVC solution,
equivalently $b_{+} = w - c \alpha ^2 / 2$,
and the uncovered backbone
is simply the set of the leaves.
We have
\begin{eqnarray}
b_{+}&&
= 1 - \sum _{k = 0}^{\infty} P(k) (1 - \alpha)^{k} - c \alpha ^2, \nonumber \\
b_{-}
&&
= \alpha, \nonumber \\
&&
\ \textrm{s.t.} \ 1 - \alpha - \beta = 0.
\end{eqnarray}
%

To compare our framework of MVC sizes with simulation on graph instances,
we adopt a hybrid algorithm
combining the GLR procedure
and the belief propagation-guided decimation (BPD) algorithm
\cite{
Weigt.Zhou-PRE-2006,
Zhao.Zhou-CPB-2014}.
Details of the BPD algorithm is left in appendix B.
The hybrid algorithm goes as:
(1) on a given graph,
the GLR procedure is applied and leaves a core;
(2) the BPD algorithm is further applied on the core,
during which
a small number of vertices with the highest marginal probabilities to be in MVCs are decimated
along with those edges attached to them;
(3) the two steps are carried out iteratively
until all edges are removed,
and finally the roots from the GLR procedure and the decimated vertices from the BPD algorithm
constitute an approximate MVC for the graph. 
We can see that on graphs without cores,
the hybrid algorithm reduces to the GLR procedure.

\subsection{MM}

For the MM problem on a graph,
we define $y$ as the relative size of a MM
(the size of edges in a MM normalized by the graph vertex size).
With the GLR procedure and the perfect matching of cores
\cite{
Karp.Sipser-IEEFoCS-1981,
Bauer.Golinelli-EPJB-2001},
on large graphs we have
\begin{eqnarray}
\label{eq:mm_short}
y = w + \frac {n}{2}.
\end{eqnarray}
Equations (\ref{eq:alpha}), (\ref{eq:beta}), (\ref{eq:n}), (\ref{eq:w}), and (\ref{eq:mm_short})
constitute our analytical framework to estimate MM sizes.
Inserting equations (\ref{eq:n}) and (\ref{eq:w}) into equation (\ref{eq:mm_short}),
we have equivalently
\begin{eqnarray}
\label{eq:mm}
y = 1 - \frac{1}{2} \sum _{k = 0}^{\infty} P(k) [(1 - \alpha)^k + \beta ^{k}] - \frac {1}{2} c \alpha (1 - \beta).
\end{eqnarray}
In the paper
\cite{Zdeborova.Mezard-JStatMech-2006}
(equations (35) - (38) therein),
from a different framework with a cavity method at the replica symmetric regime at zero temperature,
a similar result is derived by substituting
$p_{1} \rightarrow \alpha$,
$p_{2} \rightarrow \beta$, and
$\epsilon _{0} \rightarrow 1 - 2 y$.
Our contribution here,
and also in our framework of the MVC sizes,
is that,
not following the standard procedure to consider an optimization problem
as a statistical physical system to seek its ground-state energy,
we base our theory on a clear geometrical interpretation of the GLR procedure,
and our theory is much more simplified.
To test our framework of MMs on graph instances,
we adopt the Karp-Sipser algorithm
\cite{Karp.Sipser-IEEFoCS-1981}.

\section{Results}
\label{sec:results}

\begin{figure}
\begin{center}
 \includegraphics[width = 0.95 \linewidth]{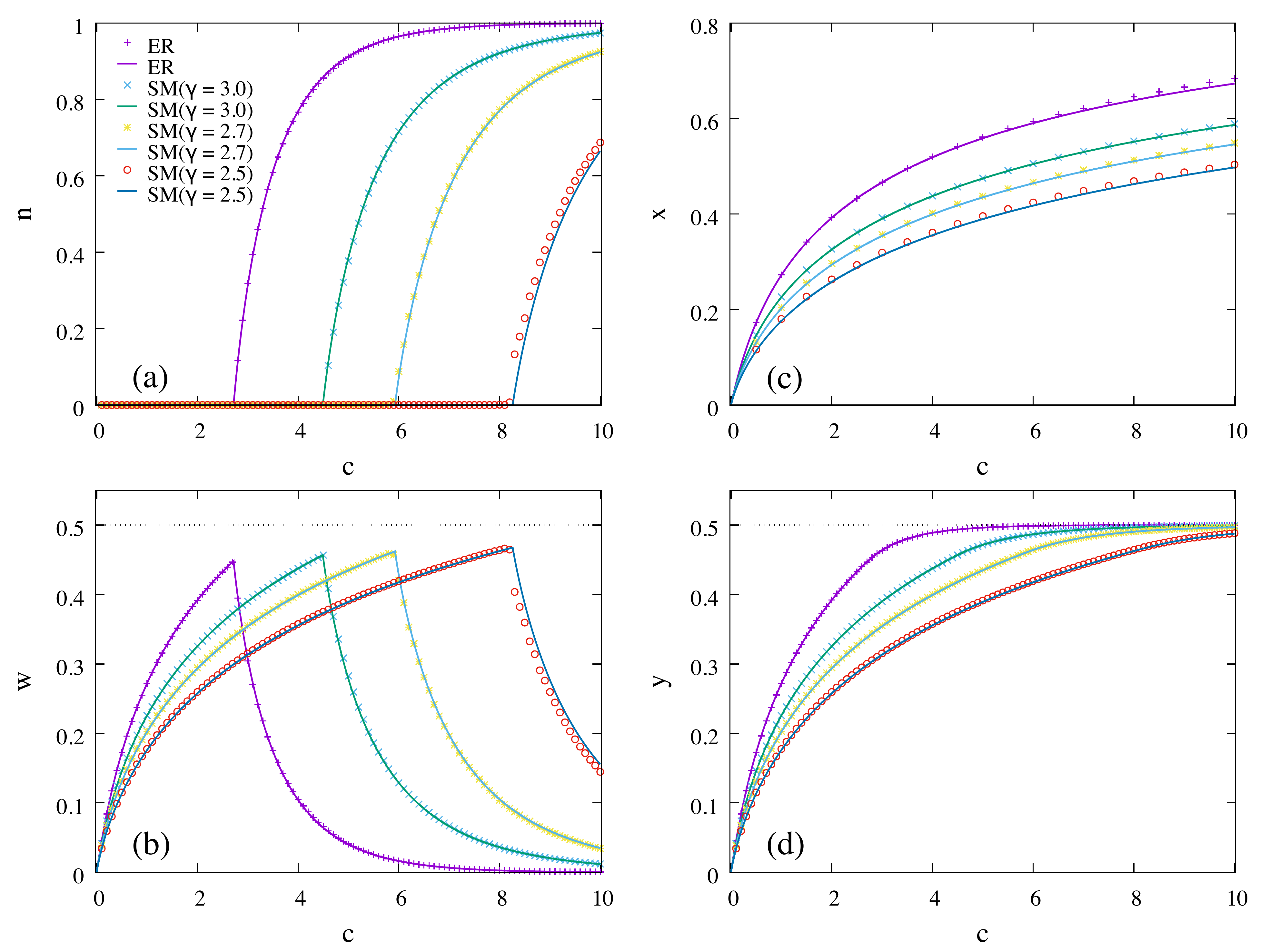}
\end{center}
\caption{
 \label{fig:glr_er_sm}
The fractions of core vertices $n$ and roots $w$ from the GLR procedure,
the MVC fraction $x$,
and the MM fraction $y$
on the Erd\"{o}s-R\'{e}nyi (ER) random graphs
and the scale-free networks generated with the static model (SM)
with a degree exponent $\gamma = 3.0, 2.7, 2.5$.
In (c), each sign is a result of the hybrid (GLR+BPD) algorithm
on a single graph instance with a vertex size $N = 10^5$;
elsewhere,  each sign is a simulation result
on a single graph instance with a vertex size $N = 10^{6}$.
Each solid line is a mean-field prediction on infinitely large graphs.}
\end{figure}
\begin{figure}
\begin{center}
 \includegraphics[width = 0.95 \linewidth]{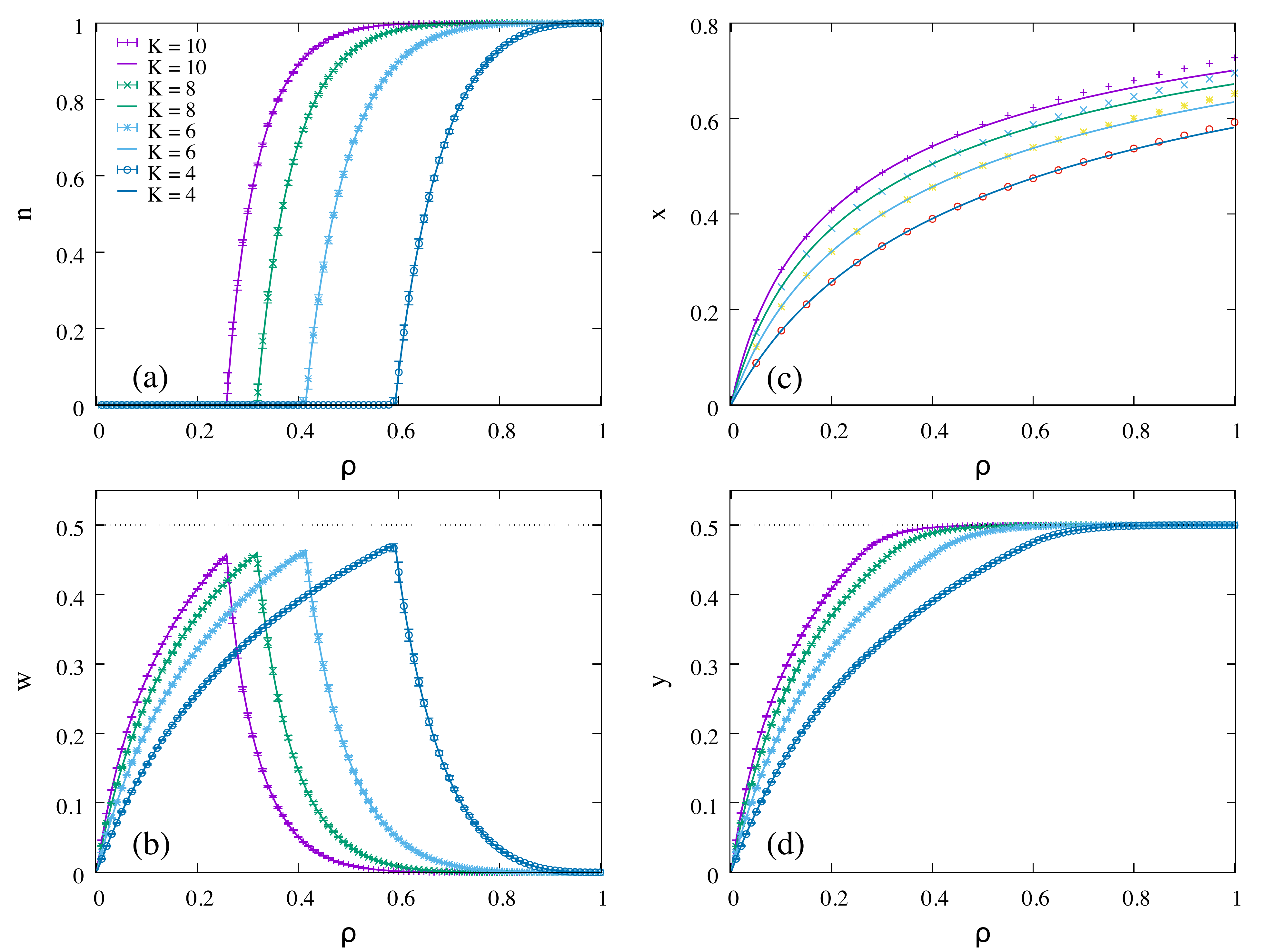}
\end{center}
\caption{
 \label{fig:glr_rr}
The fractions of core vertices $n$ and roots $w$ from the GLR procedure,
the MVC fraction $x$,
and the MM fraction $y$
on the diluted regular random (dRR) graphs with an initial degree $K = 10, 8, 6, 4$.
In (c), each sign is a result of the hybrid (GLR+BPD) algorithm
on a single diluted graph instance with a vertex size $N = 10^5$;
elsewhere, each sign is a simulation result
averaged on $40$ independently generated graph instances with a vertex size $N = 10^{5}$,
with which the standard deviation for each data point is also shown.
Each solid line is a mean-field prediction on infinitely large graphs.}
\end{figure}
\begin{figure}
\begin{center}
 \includegraphics[width = 0.95 \linewidth]{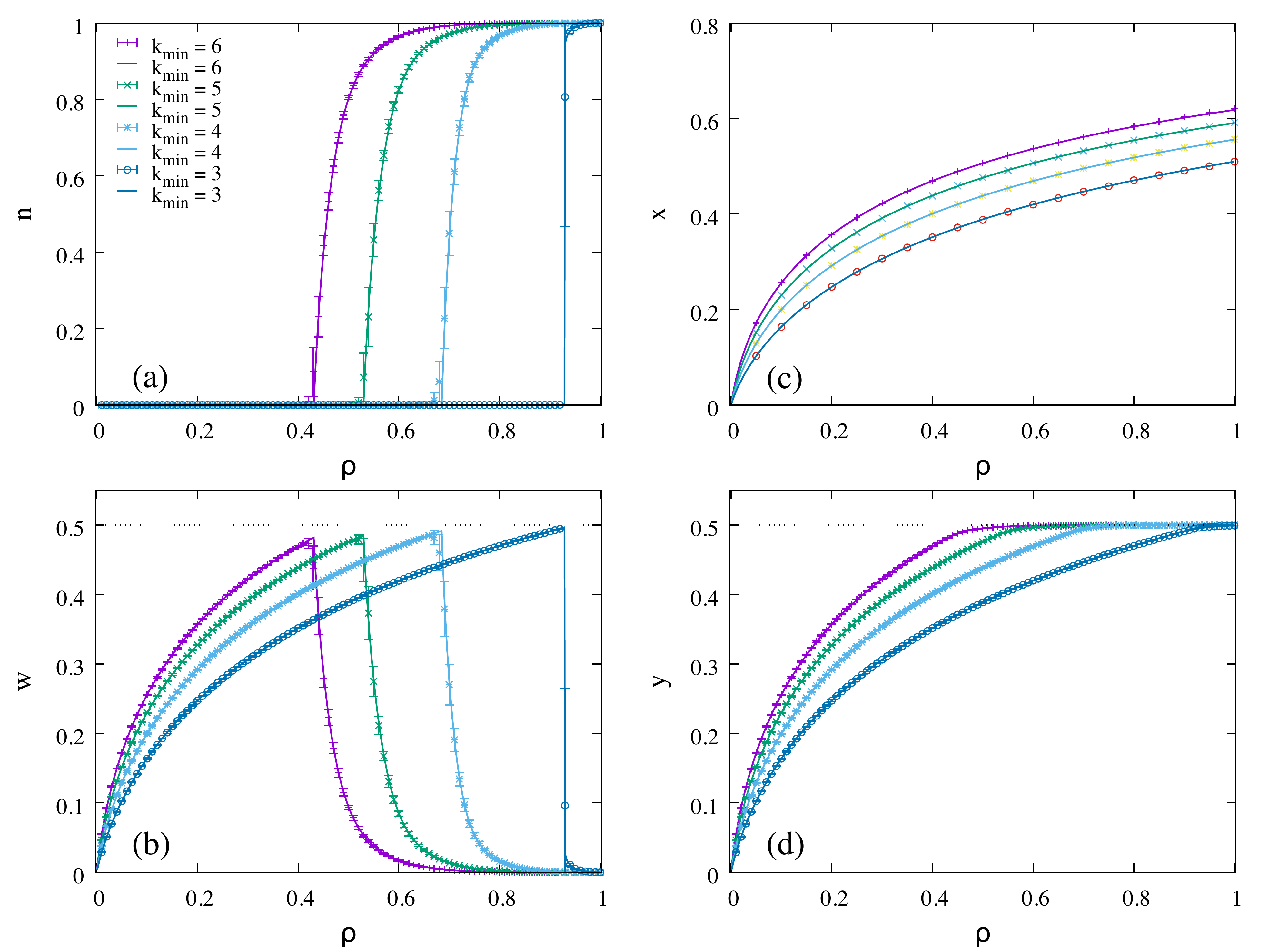}
\end{center}
\caption{
 \label{fig:glr_sf}
The fractions of core vertices $n$ and roots $w$ from the GLR procedure,
the MVC fraction $x$,
and the MM fraction $y$
on the diluted instances from a scale-free (SF) network
generated with the configurational model
with a vertex size $N = 10^{5}$,
a degree exponent $\gamma = 2.5$,
a maximal degree $k_{\rm{max}} = \sqrt N$,
and a minimal degree $k_{\rm{min}} = 6, 5, 4, 3$.
In (c), each sign is a result of the hybrid (GLR+BPD) algorithm
on a single diluted instance of the SF network;
elsewhere, each sign is a simulation result
averaged on $40$ independently generated diluted instances
of the SF network,
with which the standard deviation for each data point is also shown.
Each solid line is a mean-field prediction
with the empirical degree distribution of the SF network as input.}
\end{figure}
\begin{figure}
\begin{center}
 \includegraphics[width = 0.85 \linewidth]{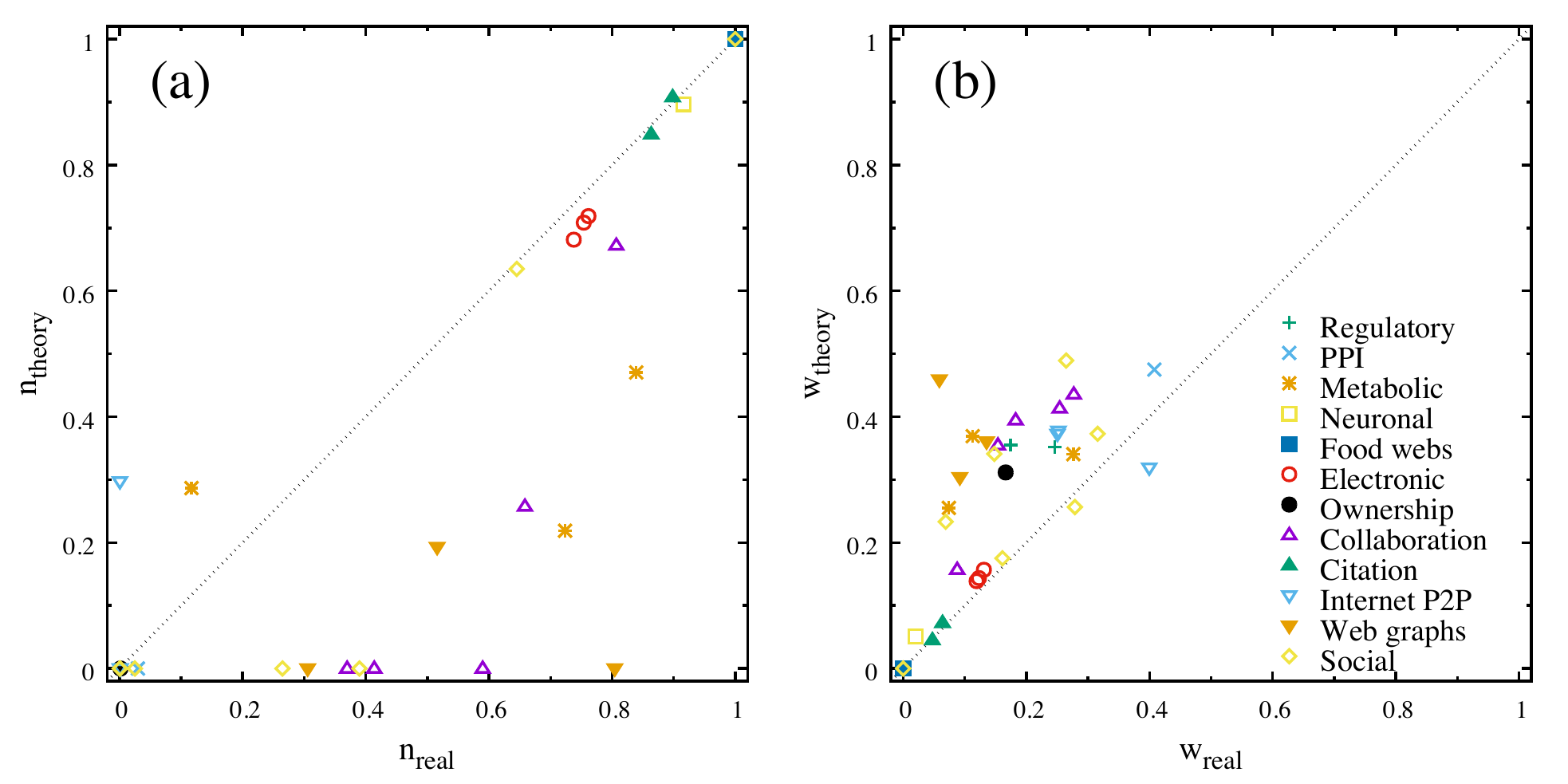}
\end{center}
\caption{
 \label{fig:glr_real}
The fractions of core vertices $n$ and roots $w$ from the GLR procedure
on real-world network instances.
The fractions $n_{\rm{real}}$ and $w_{\rm{real}}$
are directly read from the GLR procedure on the network instances.
The fractions $n_{\rm{theory}}$ and $w_{\rm{theory}}$
are the mean-field predictions
with the empirical degree distribution of each network instance as input.
Data points are grouped by network categories.}
\end{figure}

Here we apply our frameworks
for the GLR procedure, the MVCs, and the MMs
on some model random graphs and real-world networks.
Details for graph construction and equations for random graph models are left in appendix C.
A brief description of the real-world network dataset is left in appendix D.

An Erd\"os-R\'enyi (ER) random graph
\cite{
Erdos.Renyi-PublMath-1959,
Erdos.Renyi-Hungary-1960}
with a mean degree $c$
has a Poisson degree distribution as
$P(k) = e ^{- c} c ^{k} / k!$.
Results of simulation and analytical theory of the GLR procedure
are in figure \ref{fig:glr_er_sm}.
There are two regimes
in which $n$ and $w$ follow consistent scenarios respectively,
separated by the core birth point
\cite{
Bauer.Golinelli-EPJB-2001,
Liu.Csoka.Zhou.Posfai-PRL-2012}.
In the coreless regime $c < e \approx 2.71828$,
we have
$1 - \alpha - \beta = 0$,
$n = 0$,
and $w$ continuously increasing from $0$ to a maximum $w^{\star} \approx 0.44816$.
In the core regime $c > e$,
we have
$1 - \alpha - \beta > 0$,
$n$ continuously increasing from $0$,
and $w$ continuously decreasing from $w^{\star}$.
Here we present an intuitive understanding of these regimes.
When $c$ is rather small,
a graph is simply a collection of trees and has no macroscopic GCC
\cite{Molloy.Reed-RandStructAlgo-1995}.
The GLR procedure can remove all edges with few generated leaves.
After the GCC forms and with the initial leaf size ($P(1)$) decreasing,
the iterative GLR procedure relies more and more on generated leaves.
Yet due to the relative low density of edges in the graph,
the GLR procedure can still remove all edges,
resulting in a trivial $n$.
Meanwhile more single GLR steps are needed with an increasing $c$,
leading to an increasing root size $w$.
After a core forms ($c > e$) and with more edges added,
the initial leaf size further decreases.
More severely,
new leaves are difficult to generate in the procedure
due to the increasing density of edges.
Thus in general the size of GLR steps decreases,
leading to both an increasing $n$ and a decreasing $w$,
trending as $n \rightarrow 1$ and $w \rightarrow 0$
at the same time.
We will see that this picture holds for other model random graphs.

We consider the MVCs on ER random graphs.
On the ER random graphs without cores ($c \le e$),
the MVC sizes
have been derived exactly with a numeration method for the GLR procedure
(basically a discrete approach to the core percolation theory)
\cite{
Karp.Sipser-IEEFoCS-1981,
Bauer.Golinelli-EPJB-2001}.
On general ER random graphs,
the MVC sizes are estimated by the replica symmetric calculation at zero temperature
\cite{Weigt.Hartmann-PRL-2000}
and the cavity method based on the long-range frustration among MVC solutions
\cite{Zhou-PRL-2005},
both of which reduce to the exact results in the regime of graphs without cores
and serve as an approximation on graphs with cores.
The well-known replica symmetric result in
\cite{Weigt.Hartmann-PRL-2000}
states that the MVC fractions follow
\begin{eqnarray}
\label{eq:mvc_wh}
\hat{x} = 1 - \frac {1}{2c} (2 W(c) + W(c)^2),
\end{eqnarray}
in which $W(c)$ is the Lambert-W-function as $W(c) e ^{W(c)} = c$.
From our framework and equations in appendix C,
we have
\begin{eqnarray} 
x = 1 - e^{-c \alpha} - \frac {1}{2}c \alpha ^2,
\  \textrm{s.t.} \ 1 - \alpha - \beta = 0,
\end{eqnarray}
in which $\alpha$ and $\beta$ are calculated from equations (\ref{eq:alpha_er}) and (\ref{eq:beta_er}).
Here we prove that our framework simply retrieves equation (\ref{eq:mvc_wh}).
We define two auxiliary functions as
$A \equiv c e^{-c \alpha}$
and $B \equiv c \alpha$,
thus we have $x = 1 - (2A + B^{2}) / 2c$.
A list of equivalence can be carried out as
\begin{eqnarray}
A
\label{eq:mvc_A}
&&
\equiv
c e^{-c \alpha} = c e^{- c (1 - \beta)} = c \alpha \equiv B, \\
B e^{B}
\label{eq:mvc_B}
&&
\equiv
c \alpha e ^{c\alpha} = c e^{- c (1 - \beta)} e ^{c \alpha} = c e ^{- c (1 - \alpha - \beta)} = c,
\end{eqnarray}
employing only the trivial core condition and equation (\ref{eq:alpha_er}).
Thus $A = B = W(c)$,
proving that $x$ and $\hat{x}$ are the same.
This equivalence shows that,
the trivial unstable branch of fixed solutions for a percolation problem,
which is largely neglected in the present percolation study,
indeed has a physical interpretation,
which is closely related to a replica symmetric result of the underlying optimization problems.

We also consider the MMs on ER random graphs.
An enumeration method for the GLR procedure
\cite{
Karp.Sipser-IEEFoCS-1981,
Bauer.Golinelli-EPJB-2001}
and the cavity method at the first replica symmetric breaking (1RSB) level
\cite{
Zhou.OuYang-arxiv-2003}
both estimate the MM fraction as
\begin{eqnarray}
\label{eq:mm_er}
\hat{y} = \frac{1}{2} (1 - \alpha + \beta - c \alpha + c \alpha \beta),
\end{eqnarray}
in which $\alpha$ and $\beta$ are derived from
$\alpha = e^{- c e^{- c \alpha}}, \beta   = 1 - e^{- c \alpha}$.
Here we prove that our framework simply retrieves equation (\ref{eq:mm_er}).
First, it is easy to see that the two iterative equations for equation (\ref{eq:mm_er})
are just equations (\ref{eq:alpha_er}) and (\ref{eq:beta_er}).
Then, after we simplify equation (\ref{eq:mm}) with Poisson degree distributions
and remove the exponential terms
with equation (\ref{eq:alpha_er}) and (\ref{eq:beta_er}),
we finally have equation (\ref{eq:mm_er}).

We also consider our framework on the diluted regular random (dRR) graphs.
A dRR graph is obtained by an edge dilution scheme
on a regular random (RR) graph with a uniform integer degree $K$ ($\ge 2$) for each vertex,
in which a fraction $1 - \rho \in [0, 1]$ of edges
is randomly chosen and removed.
Results are shown in figure \ref{fig:glr_rr}.

We further consider the scale-free (SF) networks
\cite{Barabasi.Albert-Science-1999}
with a power-law degree distribution
$P(k) \propto k^{- \gamma}$
with a degree exponent $\gamma$.
There are several models to generate SF network instances.
First we consider the configurational model
\cite{
Newman.Strogatz.Watts-PRE-2001,
Zhou.Lipowsky-PNAS-2005},
which generates a graph instance directly from its empirical degree distribution.
Results on diluted SF networks with $\gamma = 2.5$ are in shown figure \ref{fig:glr_sf}.
Theory and simulation correspond very well except those results of $n$ and $w$ close to core birth.
An explanation is the finite-size effect
at the phase transition points on the relatively small graph instances.
We then consider the asymptotical SF networks
generated with the static model
\cite{
Goh.Kahng.Kim-PRL-2001,
Catanzaro.PastorSatorras-EPJB-2009}.
Results are shown in figure \ref{fig:glr_er_sm}.
For $n$ and $w$,
a discernible discrepancy sets in with small $\gamma$.
It has been analytically proved
\cite{
Goh.Kahng.Kim-PRL-2001,
Catanzaro.PastorSatorras-EPJB-2009}
that the SF networks generated with the static model with $\gamma > 3$
behave more like ER random graphs with $\gamma$ increasing,
while those with $\gamma < 3$
show more significant degree-degree correlation with $\gamma$ decreasing.
Besides the degree-degree correlation in the SF networks with $\gamma < 3$,
the loop structure is also a contributing factor to the deviation of the mean-field results.
An analytical discussion of finite size correction through loops can be found in
\cite{Lucibello.etal-PRE-2014}.

We also test the GLR procedure on real-world networks.
Results from simulation and theory are shown in figure \ref{fig:glr_real}.
For most data points, we can see two straight-forward tendencies.
The first one is that the discrepancies between simulation and theoretical results show a negative tendency.
It is easy to understand,
because cores and roots are much like complementary structure of a graph.
The second tendency is that large discrepancies between simulation and theory
exist among most of the network instances.
It shows a fundamental limitation of the basis of our analytical theory,
the local tree-structure assumption
\cite{Mezard.Montanari-2009}.
Real networks show rich mesoscopic and higher-order structures at various scales,
such as the degree-degree correlation among neighboring vertices
\cite{Newman-PRL-2002},
the community structure and the modularity
\cite{Newman-NatPhys-2012},
and the hierarchical organizations
\cite{CorominasMurtra.etal-PNAS-2013}.
All these factors can be slimly captured by the degree distribution,
which is the only information of graph structure into our mean-field theory,
thus can fail the analytical prediction.

\section{Conclusion}
\label{sec:conclusion}

The geometrical implication of roots and cores from the GLR procedure
in the MVC and the MM problems
has long been formulated in an algorithmic sense,
yet on the analytical side a thorough exploration is still missing
except the mean-field theory of core emergence.
In this paper,
by completing the missing theory of roots from the GLR procedure,
we develop a unified framework to estimate MVC and MM sizes
on random graphs with or without cores:
with the branch of trivial fixed solutions of the iterative cavity equations of the GLR procedure,
we derive the zero-temperature replica symmetric solution of the MVC fractions;
with the branch of stable fixed solutions,
we reproduce the zero-temperature replica symmetric estimation of the MM fractions.
Our frameworks are built on a geometrical interpretation of roots and cores,
which makes them different from the typical formulation of optimization problems
into statistical physical systems with zero or finite temperatures.
Besides, our frameworks have a simple and easily interpretable form,
without involving complicated statistical physical tricks.

There are still points deserving further exploration,
particularly the physical interpretation of the unstable branch of self-consistent equations
common in mean-field theories for random systems.
A recent study
\cite{Zhou-PRL-2019}
in the Bethe-Peierls mean-field theory for the $Q$-state Potts model on random graphs
discusses the physical role of the middle unstable branch of cavity probabilities
in the microcanonical discontinuous phase transitions.

\section{Acknowledgements}

This work is supported by
the National Natural Science Foundation of China (Grant Nos. 11421063 and 11747601)
and the Chinese Academy of Sciences (Grant No. QYZDJ-SSW-SYS018).
J-H Zhao is supported partially by
the Key Research Program of Frontier Sciences of the Chinese Academy of Sciences (Grant No. QYZDB-SSW-SYS032) and
the National Natural Science Foundation of China (Grant No. 11605288).
J-H Zhao thanks Prof. Pan Zhang (ITP-CAS) for his hospitality.

\section{Appendix A: A discrete theoretical formulation of the GLR procedure}

\subsection{A discrete form of $\alpha$ and $\beta$}

We consider the discretized GLR procedure on an undirected graph $G = \{V, E\}$.
At each $t$-th time-step with a discrete index $t \ge 0$,
all leaves and their roots are determined,
and their adjacent edges are removed.
Beware that a 'time-step' of GLR procedure here
usually involves multiple 'single steps' of the GLR procedure
in which only one root is removed at a step.
We can calculate the pertinent quantities just after each $t$-th time-step.
In the case of infinite time steps as $t \rightarrow \infty$,
the discretized GLR procedure is iterated until there is no leaf left.
Correspondingly,
the residual subgraph constitutes the core,
and the roots can be counted by summing those removed during all the time-steps.

Following a similar logic of the definition of
$\alpha$ and $\beta$ in equations (\ref{eq:alpha}) and (\ref{eq:beta}),
we define cavity probabilities $\alpha ^{(t)}$ and $\beta ^{(t)}$ for the $t$-th time-step ($t \ge 0$).
With a randomly chosen edge $(i, j) \in G$ and following the vertex $i$ to the vertex $j$,
we define
$\alpha ^{(t)}$ ($\beta ^{(t)}$)
as the probability that $j$ becomes a leaf (a root)
at exactly the $t$-th time-step on $G \backslash (i, j)$.
We also define the cumulative cavity probabilities
$\alpha ^{(c, t)} \equiv \sum _{t' = 0}^{t} \alpha ^{(t')}$
($\beta ^{(c, t)} \equiv \sum _{t' = 0}^{t} \beta ^{(t')}$)
as the probability that $j$ becomes a leaf (a root)
before or at the $t$-th time-step on $G \backslash (i, j)$.
We then derive the iterative equations of $\alpha ^{(t)}$ and $\beta ^{(t)}$ with $t \ge 0$
based on the local tree-structure approximation on sparse random graphs.
\begin{eqnarray}
\alpha ^{(t)}
&&
=
\left\{
\begin{array}{lcr}
Q(1),
& \mbox{for}
& t = 0; \\
\sum _{k = 1}^{\infty} Q(k) (\beta ^{(c, 0)})^{k - 1} - Q(1),
& \mbox{for}
& t = 1; \\
\sum _{k = 1}^{\infty}
Q(k) [(\beta ^{(c, t - 1)})^{k - 1}
- (\beta ^{(c, t - 2)})^{k - 1}],
& \mbox{for}
& t \ge 2;
\end{array}
\right. \\
\beta ^{(t)}
&&
=
\left\{
\begin{array}{lcr}
\sum _{k = 1}^{\infty} Q(k) [1 - (1 - \alpha ^{(c, 0)})^{k - 1}],
& \mbox{for}
& t = 0; \\
\sum _{k = 1}^{\infty}
Q(k) \{[1 - (1 - \alpha ^{(c, t)})^{k - 1}]
- [1 - (1 - \alpha ^{(c, t - 1)})^{k - 1}]\},
& \mbox{for}
& t \ge 1.
\end{array}
\right. 
\end{eqnarray}
Here is an explanation for the equations.
With a randomly chosen edge $(i, j) \in G$,
we consider the state of $j$
from a cavity graph $G \backslash j$
to another cavity graph $G \backslash (i, j)$
at the $t$-th time-step under the GLR procedure
after the corresponding edge addition.
For $\alpha ^{(t)}$,
if the vertex $j$ becomes a leaf at the $t$-th time-step on $G \backslash (i, j)$,
all the nearest neighbors of $j$ except $i$
should all be removed as roots before or at the $(t - 1)$-th time-step on $G \backslash j$,
among which there should be at least one nearest neighbors being roots
at the $(t - 1)$-th time-step.
For $\beta ^{(t)}$,
if the vertex $j$ becomes a root at the $t$-th time-step on $G \backslash (i, j)$,
among the nearest neighbors of $j$ except $i$
there should be at least one nearest neighbors being leaves
before or at the $t$-th time-step on $G \backslash j$,
among which there should be at least one nearest neighbors being leaves at the $t$-th time-step.

We can further derive iterative equations for
$\alpha ^{(c,t)}$ and $\beta ^{(c, t)}$ with $t \ge 0$.
For $\alpha ^{(c, t)}$,
it's simple to verify that
$\alpha ^{(c, 0)} \equiv \alpha ^{(0)} = Q(1)$, and
$\alpha ^{(c, 1)} \equiv \alpha ^{(0)} + \alpha ^{(1)} = \sum _{k = 1}^{\infty} Q(k) (\beta ^{(c, 0)})^{k - 1}$.
For $t \ge 2$, we have
\begin{eqnarray}
\alpha ^{(c, t)}
&&
\equiv \alpha ^{(c, 1)} + \alpha ^{(2)} + ... + \alpha ^{(t)}  \nonumber \\
&&
= \alpha ^{(c, 1)}
+ \sum _{k = 1}^{\infty}
Q(k) (\beta ^{(c, 1)})^{k - 1} \nonumber
- \sum _{k = 1}^{\infty}
Q(k) (\beta ^{(c, 0)})^{k - 1} + ... \nonumber \\
&&
+ \sum _{k = 1}^{\infty} Q(k) (\beta ^{(c, t - 1)})^{k - 1}
-  \sum _{k = 1}^{\infty} Q(k) (\beta ^{(c, t - 2)})^{k - 1} \nonumber \\
&&
= \sum _{k = 1}^{\infty}
Q(k) (\beta ^{(c, t - 1)})^{k - 1}. \nonumber
\end{eqnarray}
We can see that the above equation also holds for $t = 1$.
For $\beta ^{(c, t)}$,
we have $\beta ^{(c, 0)} \equiv \beta ^{(0)}$.
With $t \ge 1$,
we have
\begin{eqnarray}
\beta ^{(c, t)}
&&
\equiv \beta ^{(0)} + \beta ^{(1)} + ... + \beta ^{(t)} \nonumber \\
&&
= \beta ^{(0)}
+ \sum _{k = 1}^{\infty}
Q(k) [1 - (1 - \alpha ^{(c, 1)})^{k - 1}]
- \sum _{k = 1}^{\infty}
Q(k) [1 - (1 - \alpha ^{(c, 0)})^{k - 1}] +...  \nonumber \\
&&
+ \sum _{k = 1}^{\infty}
Q(k) [1 - (1 - \alpha ^{(c, t)})^{k - 1}]
- \sum _{k = 1}^{\infty}
Q(k) [1 - (1 - \alpha ^{(c, t - 1)})^{k - 1}] \nonumber \\
&&
= \sum _{k = 1}^{\infty}
Q(k) [1 - (1 - \alpha ^{(c, t)})^{k - 1}]. \nonumber
\end{eqnarray}
The above equation also holds for $t = 0$.

Summing the above results,
$\alpha ^{(c, t)}$ and $\beta ^{(c, t)}$ with $t \ge 0$
can be derived from the initial condition $\alpha ^{(c, 0)} = Q(1)$
and the updating equations
\begin{eqnarray}
\alpha ^{(c, t)}
\label{eq:alpha_ct_discrete}
&&
=
\sum _{k = 1}^{\infty}
Q(k) (\beta ^{(c, t - 1)})^{k - 1}, \\
\beta ^{(c, t)}
\label{eq:beta_ct_discrete}
&&
=
1 - \sum _{k = 1}^{\infty}
Q(k) (1 - \alpha ^{(c, t )})^{k - 1}.
\end{eqnarray}
%

We present a numerical method to calculate
the cavity probabilities $(\alpha ^{(c,t)}, \beta ^{(c, t)})$,
thus $(\alpha ^{(t)}, \beta ^{(t)})$,
with any $t \ge 0$.
With the initial condition $\alpha ^{(c, 0)} = Q(1)$,
we can derive $\beta ^{(c, 0)}$
with equation (\ref{eq:beta_ct_discrete}),
thus we have $(\alpha ^{(c, 0)}, \beta ^{(c, 0)})$.
With any cavity probability $(\alpha ^{(c, t')}, \beta ^{(c, t')})$ with $t' \ge 0$,
we can calculate $\alpha ^{(c, t' + 1)}$ with equation (\ref{eq:alpha_ct_discrete})
and $\beta ^{(c, t' + 1)}$ with equation (\ref{eq:beta_ct_discrete}),
thus we have
$(\alpha ^{(c, t' + 1)}, \beta ^{(c, t' + 1)})$.
In such a progressive process, we can calculate
$(\alpha ^{(c,t)}, \beta ^{(c, t)})$
with any $t \ge 0$.
To stop the probability updating as 
$(\alpha ^{(c,t)}, \beta ^{(c, t)})$ converges at large $t$,
we can set a simple criterion:
we terminate at the $T$-th step
once $|\alpha ^{(c, T)} - \alpha ^{(c, T - 1)}| + |\beta ^{(c, T)} - \beta ^{(c, T - 1)}| < \epsilon$,
in which $\epsilon$ is a small positive number such as $\epsilon = 10^{- 8}$.

\subsection{A discrete form of $n$, $l$, and $w$}

Here we calculate some quantities
just after the $t$-th time-step of the discretized GLR procedure.
In this context,
the relative sizes of vertices and edges
in the residual subgraph and
the accumulated removed roots
(all normalized by the vertex size)
are denoted as 
$n ^{(c, t)}$,
$l^{(c, t)}$,
and $w ^{(c, t)}$
with $t \ge 0$,
respectively.

For $n ^{(c, t)}$,
it can be written down as
\begin{eqnarray}
\label{eq:n_discrete}
n ^{(c, t)} 
&&
=
\sum _{k  = 2}^{\infty} P(k)
\sum _{s = 2}^{k}
 \left(\begin{array}{c} k \\ s \end{array}\right)
(\beta ^{(c, t)})^{k -s}
(1 - \alpha ^{(c, t)} - \beta ^{(c, t)})^{s}.
\end{eqnarray}
Here is a simple explanation.
With a randomly chosen vertex $i$,
we move from a cavity graph $G \backslash i$ to the original graph $G$
after some edge addition.
If $i$ is in the residual subgraph of $G$ just after the $t$-th time-step,
on $G \backslash i$ it should have:
(1) no nearest neighbor turning into a leaf before or at the $t$-th time-step;
(2) at least two nearest neighbors
which are also in the residual subgraph just after the $t$-th time-step; and
(3) all the other nearest neighbors becoming roots before or at the $t$-th time-step.

For $l ^{(c, t)}$, we have
\begin{eqnarray}
\label{eq:l_discrete}
l ^{(c, t)} 
&&
=
\frac {1}{2}
c (1 - \alpha ^{(c, t)} - \beta ^{(c, t)})^{2}.
\end{eqnarray}
Here is a simple explanation.
We move from a cavity graph $G \backslash (i, j)$ to $G$
as $(i, j) \in G$ is a randomly chosen edge.
If $(i, j)$ is in the residual graph of $G$ just after the $t$-th time-step,
both its end-vertices should be in the residual graph of
$G \backslash (i, j)$ just after the $t$-th time-step,
thus we have the squared term.
The multiplicative factor $c / 2 (= |E| / |V|)$ is the relative size of edges to vertices.

For $w^{(c, t)}$,
we follow a similar calculation
in the context of a generalized GLR procedure
for the minimum dominating set problem
\cite{
Zhao.Zhou-JStatPhys-2015,
Zhao.Zhou-LNCS-2015}.
For the GLR procedure here,
we have
\begin{eqnarray}
\label{eq:w_discrete}
w ^{(c, t)}
&& =
\frac {1}{2} P(1) Q(1)
+ \sum _{k = 2}^{\infty}
P(k) [1 - (1 - \alpha ^{(c, t)})^{k}] \nonumber \\
&&
- \frac {1}{2} \sum _{k = 2}^{\infty} P(k) k \alpha ^{(1)} (\beta ^{(0)})^{k - 1} \nonumber \\
&&
- \frac {1}{2} \sum _{\tau = 2}^{t} \sum _{k = 2}^{\infty}
P(k) k \alpha ^{(\tau)} [(\beta ^{(c, \tau - 1)})^{k - 1} - (\beta ^{(c, \tau - 2)})^{k - 1}] \nonumber \\
&&
- \sum _{\tau = 1}^{t - 1} \sum _{k = 2}^{\infty}
P(k) k \alpha ^{(\tau + 1)} (\beta ^{(c, \tau - 1)})^{k - 1}.
\end{eqnarray}
Here is a simple explanation.
(1) The first term counts the isolated edge
with both end-vertices as leaves.
(2) The second term is a summation
of the possibilities that a random vertex, say $i$,
having some nearest neighbor being leaves
before or at the $t$-th time-step
on the cavity graph $G \backslash i$.
Yet there are recounting terms from the newly generated isolated edges
and terms contradicting the pruning process of the discretized GLR procedure,
which are further subtracted in the following three terms.
(3) - (4) The recounting case happens
when a certain edge turns out to be an isolated edge on $G$ at certain time step.
Specifically,
at the $\tau$-th $(1 \le \tau \le t)$ time-step of the GLR procedure on $G \backslash j$,
if all the nearest neighbors of $j$ except $i$
are pruned already as roots before or at the $(\tau - 1)$-th time-step
among which some nearest neighbors are pruned at exactly the $(\tau - 1)$-th time-step,
$j$ will become a leaf at the $\tau$-th time-step on $G \backslash (i, j)$.
If $i$ happens to be a leaf at the $\tau$-th time-step on $G \backslash (i, j)$,
an isolated edge $(i, j)$ emerges on $G$.
The third and fourth terms sums these probabilities
of the case  of $\tau = 1$ and $2 \le \tau \le t$ respectively.
(5) The fifth term deals with the contradiction case
when a product of cavity probabilities from the second term
does not correspond to a proper state transition in the discretized GLR procedure.
At the $\tau$-th time-step ($1 \le \tau \le t - 1$) of the GLR procedure on $G \backslash j$,
if all the nearest neighbors of $j$ except $i$
are pruned as roots before or at the $(\tau - 1)$-th time-step,
$j$ will be a leaf before or at the $\tau$-th time-step
on $G \backslash (i, j)$.
A contradiction happens when $i$ turns into a leaf
at the $(\tau + 1)$-th time-step on $G \backslash (i, j)$.
The reason is that,
in a proper scenario, $i$ should be a leaf or simply a root at the $\tau$-th time-step on $G \backslash (i, j)$,
thus the states of $i$ and $j$ can be well defined in $G$.

Below we further simplify $w ^{(c, t)}$.
On the RHS of equation (\ref{eq:w_discrete}),
in the second term
we reformulate $\sum _{k = 2}^{\infty}$ with $\sum _{k = 0}^{\infty}$,
in the fifth term
we change from $\sum _{\tau = 1}^{t - 1}$ to $\sum _{\tau = 2}^{t}$,
and we further combine the fourth and the fifth terms.
We also see $\alpha ^{(0)} = Q(1)$.
We then have
\begin{eqnarray}
w ^{(c, t)}
&& =
\frac {1}{2} P(1) \alpha ^{(0)}
+ \sum _{k = 0}^{\infty}
P(k) [1 - (1 - \alpha ^{(c, t)})^{k}]
- P(1) \alpha ^{(c, t)} \nonumber \\
&&
- \frac {1}{2} \sum _{k = 2}^{\infty} P(k) k \alpha ^{(1)} (\beta ^{(0)})^{k - 1} \nonumber \\
&&
- \frac {1}{2} \sum _{\tau = 2}^{t} \sum _{k = 2}^{\infty}
P(k) k \alpha ^{(\tau)} [(\beta ^{(c, \tau - 1)})^{k - 1} + (\beta ^{(c, \tau - 2)})^{k - 1}]. \nonumber
\end{eqnarray}
We further move from $\sum _{k = 2}^{\infty}$ to $\sum _{k = 1}^{\infty}$ in the last two terms,
and rearrange the equation.
\begin{eqnarray}
w ^{(c, t)}
&& = 
\sum _{k = 0}^{\infty} P(k) [1 - (1 - \alpha ^{(c, t)}) ^{k}]
+ \frac {1}{2} P(1) \alpha ^{(0)}
- P(1) \alpha ^{(c, t)} \nonumber \\
&&
- \frac {1}{2} \sum _{k = 1}^{\infty}
P(k) k \alpha ^{(1)} (\beta ^{(0)})^{k - 1}
+ \frac {1}{2} P(1) \alpha ^{(1)} \nonumber \\
&&
- \frac {1}{2} \sum _{\tau = 2}^{t} \sum _{k = 1}^{\infty}
P(k) k \alpha ^{(\tau)} [(\beta ^{(c, \tau - 1)})^{k - 1} + (\beta ^{(c, \tau - 2)})^{k - 1}] \nonumber \\
&&
+ \frac {1}{2} \sum_{\tau = 2}^{t} P(1) \alpha ^{(\tau)} [1 + 1]. \nonumber
\end{eqnarray}
We combine all the non-summation terms with $P(1)$,
and further adopt the equivalence $c Q(k) = k P(k)$.
We have
\begin{eqnarray}
w ^{(c, t)}
&& =
\sum _{k = 0}^{\infty}
P(k) [1 - (1 - \alpha ^{(c, t)})^{k}]
- \frac {1}{2} c Q(1) \alpha ^{(c, 1)} \nonumber \\
&&
- \frac {1}{2} \sum _{k = 1}^{\infty}
c Q(k) \alpha ^{(1)} (\beta ^{(0)})^{k - 1} \nonumber \\
&&
- \frac {1}{2} \sum _{\tau = 2}^{t} \sum _{k = 1}^{\infty}
c Q(k) \alpha ^{(\tau)}
[(\beta ^{(c, \tau - 1)})^{k - 1} + (\beta ^{(c, \tau - 2)})^{k - 1}]. \nonumber
\end{eqnarray}
Considering the equivalence $\alpha ^{(0)} = Q(1)$ for the second term
and equation (\ref{eq:alpha_ct_discrete}) for the last two terms,
we have
\begin{eqnarray}
w ^{(c, t)}
&&
=
\sum _{k = 0}^{\infty} P(k) [1 - (1 - \alpha ^{(c, t)})^{k}]
- \frac {1}{2} c \alpha ^{(0)} \alpha ^{(c, 1)} \nonumber \\
&&
- \frac {1}{2} c \alpha ^{(1)} \alpha ^{(c, 1)}
- \frac {1}{2} \sum _{\tau = 2}^{t} c \alpha ^{(\tau)}
[\alpha ^{(c, \tau)} + \alpha ^{(c, \tau - 1)}]. \nonumber
\end{eqnarray}
Combining the second and third terms as $\alpha ^{(c, 1)} \equiv \alpha ^{(0)} + \alpha ^{(1)}$,
and considering the definition $\alpha ^{(t)} \equiv \alpha ^{(c, t)} - \alpha ^{(c, t - 1)}$,
we have
\begin{eqnarray}
w ^{(c, t)}
&&
= \sum _{k = 0}^{\infty} P(k) [1 - (1 - \alpha ^{(c, t)})^{k}]
-  \frac {1}{2} c (\alpha ^{(c, 1)})^2 \nonumber \\
&&
- \frac {1}{2} \sum _{\tau = 2}^{t} c [(\alpha ^{(c, \tau)})^2 - (\alpha ^{(c, \tau - 1)})^2]. \nonumber
\end{eqnarray}
Expanding the third term, we finally have
\begin{eqnarray}
w ^{(c, t)}
\label{eq:w_discrete_simplified}
= 1 - \sum _{k = 0}^{\infty} P(k) (1 - \alpha ^{(c, t)})^{k}
- \frac {1}{2} c (\alpha ^{(c, t)})^{2}.
\end{eqnarray}

\subsection{At infinite time-steps}

When $t \rightarrow \infty$,
the discrete description corresponds to
the termination of the GLR procedure on graphs.
The solution $(\alpha ^{(c, \infty)}, \beta ^{(c, \infty)})$
can be derived as the convergent values of equations
(\ref{eq:alpha_ct_discrete}) and
(\ref{eq:beta_ct_discrete}),
with which we can calculate
$n^{(c, \infty)}$,
$l^{(c, \infty)}$, and
$w^{(c, \infty)}$.
In another word,
$\alpha ^{(c, \infty)}$,
$\beta ^{(c, \infty)}$,
$n^{(c, \infty)}$,
$l^{(c, \infty)}$, and
$w^{(c, \infty)}$ reduce respectively to
$\alpha$, $\beta$, $n$, $l$, and $w$
in the main text.

\section{Appendix B: Belief propagation algorithms for the MVC problem}

Here we discuss briefly how the MVC problem can be mapped to a statistical physical system
and its approximate mean-field message-passing algorithms in the replica symmetric regime,
such as the belief propagation (BP) algorithm
and belief propagation-guided decimation (BPD) algorithm.
More detailed discussions on message-passing algorithms can be found in
\cite{
Weigt.Zhou-PRE-2006,
Zhang.Zeng.Zhou-PRE-2009}.

The MVC problem, an optimization problem in nature,
can be considered as a statistical physical system,
in which its optimum corresponds to the (zero-temperature) ground-state energy
of the corresponding physical formulation.
We discuss the problem in the context of an undirected graph $G = \{V, E\}$
with a vertex set $V$ and an edge set $E$.
For each vertex, say $i \in V$,
a covering state can be assigned as $s_{i} = 1 (0)$ as the vertex being in (not in) a MVC.
A covering state for $G$
can be denoted as a vector $\vec {s} \equiv (s_{1}, ..., s_{|V|})$.
The energy $E(\vec s)$ for a covering state $\vec s$
is simply $E(\vec s) = \sum _{i \in V} s_{i}$.
For the physical system on $G$ with a finite inverse temperature $x$,
the partition function is
\begin{eqnarray}
\label{eq:Zx}
Z(x)
= \sum _{\vec {s}}
\prod _{i \in V} e^{- x s_{i}}
\prod _{(j, k) \in E} [1 - (1 - s_{j}) (1 - s_{k})].
\end{eqnarray}
In the equation,
the summation runs on all the $2^{|V|}$ possible covering state configurations,
the first product counts all the reweighting Boltzmann factors for a covering state,
and the second product selects any covering state
as a proper VC under the topological constraint.
In all, only VC configurations among all the possible covering states
contribute to the partition function of the physical system.
From the partition function,
the thermal quantities of the physical system
can de derived with standard statistical physical methods.
In the case of $x \rightarrow \infty$,
only those VC configurations with the smallest cardinality are considered,
thus evaluating equation (\ref{eq:Zx}) reduces to the MVC problem.

Calculating equation (\ref{eq:Zx}) analytically is extremely difficult,
and the mean-field theoretical methods,
such as the replica trick and the cavity method from the spin glass theory,
are adopted.
The BP algorithm of the cavity method is based on the assumption
of locally tree-like structure on sparse random graphs,
and works in the replica symmetric regime,
in which all the solutions are assumed to be organized in a macroscopic state.
On any edge $(i, j) \in G$ from the vertex $i$ to the vertex $j$,
we define the cavity probability (or cavity message) $p_{i \rightarrow j}$
as the probability of $i$ to be in a MVC on the cavity graph $G \backslash (i, j)$.
We can see that there are in total $2 |E|$ cavity probabilities for $G$.
For any vertex $i \in G$,
we define the marginal probability $p_{i}$
as the probability of $i$ to be in a MVC on $G$.
Here we consider the case from a cavity graph $G \backslash i$ to $G$.
If the vertex $i$ is in a MVC,
a Boltzmann factor $e^{- x}$ is introduced;
otherwise,
the probability of $i$ to be not in a MVC is
$\prod _{k \in \partial i} p_{k \rightarrow i}$.
Thus we have the formulation of marginal probabilities based on the cavity probabilities as
\begin{eqnarray}
\label{eq:bp_marginal}
p _{i}
= \frac {e^{- x}}{e^{-x} + \prod _{k \in \partial i} p_{k \rightarrow i}}.
\end{eqnarray}
Following the above logic,
in the case from $G \backslash i$ to $G \backslash (i, j)$,
we have the self-consistent equation for cavity probabilities as
\begin{eqnarray}
\label{eq:bp_updating}
p _{i \rightarrow j}
= \frac {e^{- x}}{e^{-x} + \prod _{k \in \partial i \backslash j} p_{k \rightarrow i}},
\end{eqnarray}
in which $\partial i \backslash j$ is the set of the nearest neighbors of $i$ excluding $j$.
With the fixed point of the $2 |E|$ self-consistent equations of cavity probabilities on $G$
with equation (\ref{eq:bp_updating}),
the thermal properties of the statistical physical system can be estimated.
The energy density or the fraction $e$ of the MVCs is estimated
as the averaged marginal probability of any vertex.
We have
\begin{eqnarray}
e
= \frac {1}{|V|} \sum _{i \in V} p_{i}
= \frac {1}{|V|} \sum _{i \in V} \frac {e^{- x}}{e^{- x} + \prod _{j \in \partial i} p _{j \rightarrow i}}.
\end{eqnarray}
The free energy of the physical system is summed from
the free energy perturbations of adding a vertex or an edge to the corresponding cavity graph.
We have the free energy density
\begin{eqnarray}
f
= \frac {1}{|V|} ( \sum _{i \in V} \Delta F_{i} - \sum _{(i, j) \in E} \Delta F_{(i, j)} ),
\end{eqnarray}
in which $\Delta F_{i}$ for any vertex $i \in V$
is the free energy contribution by adding $i$ into the cavity graph $G \backslash i$,
and $\Delta F _{(i, j)}$ for any edge $(i, j) \in E$
is the free energy contribution by adding $(i, j)$ into the cavity graph $G \backslash (i, j)$.
We have
\begin{eqnarray}
\Delta F_{i}
&&
= - \frac {1}{x} \ln (e^{- x} + \prod _{j \in \partial i} p _{j \rightarrow i}), \\
\Delta F_{(i, j)}
&&
= - \frac {1}{x} \ln [1 - (1 - p_{i \rightarrow j})(1 - p _{j \rightarrow i})].
\end{eqnarray}
Thus we have the entropy density as $s = x (e - f)$.

The BP algorithm estimates the average physical properties in a ensemble sense,
yet the BPD algorithm applies on graph instances and approximates solution configurations.
Basically, the BPD algorithm guides the searching for approximate solutions
by fixing into certain states those vertices or edges with the most biased marginal probabilities
in an iterative graph simplification process.
The component of the BPD algorithm in the hybrid algorithm of the main text goes as:
(1) the cavity probabilities are initialized randomly $\in (0, 1)$
on each edge of the current core of the residual graph from the GLR procedure;
(2) all the cavity probabilities are updated
following equation (\ref{eq:bp_updating})
until a given maximal sweep size $N_{\rm{up}}$
or a convergence of cavity probabilities under some criterion;
(3) the marginal probability for each vertex is calculated with equation (\ref{eq:bp_marginal});
(4) in a single decimation step,
a small size $N_{\rm{d}}$ of the vertices in the core with the largest marginal probabilities
are selected into a VC
and are further removed along with their adjacent edges.
For the convergence criterion to terminate the probability updating at the $T$-th ($T \ge 1$) sweep,
we set for convenience $\max _{(i, j) \in E} | p _{i \rightarrow j}^{(T)} - p _{i \rightarrow j}^{(T - 1)} | < \epsilon$,
in which $p _{i \rightarrow j} ^{(T)}$
is the value of $p _{i \rightarrow j}$ at the $T$-th sweep of the probability updating
and the positive parameter $\epsilon \ll 1$.
For the size of removed vertices in a single decimation step,
we set $N_{\rm{d}} = \max \{N_{\rm{core}} / f_{\rm{d}}, N_{\rm{dmin}} \}$,
in which $N_{\rm{core}}$ is the vertex size of the current core,
$f_{\rm{d}}$ as a fraction coeffiecient,
and $N_{\rm{dmin}}$ as the smallest size of vertices to be removed.
In choosing a proper inverse temperature $x$ for the BPD algorithm,
we set $x$ as its largest value at which the corresponding entropy density $s$ vanishes.

In the main text,
we set $x = 10.0$,
$N_{\rm{up}} = 200$,
$\epsilon = 10^{- 8}$,
$f_{\rm{d}}= 200$,
and $N_{\rm{dmin}} = 1$.
Beware that,
by tuning $x$ adaptively for specific graph ensembles,
increasing $N_{\rm{up}}$ and $f_{\rm{d}}$,
and decreasing $\epsilon$,
we can improve results from the BPD algorithm.

\section{Appendix C: Equations for the GLR procedure on random graphs}

\subsection{Erd\"os-R\'enyi random graphs}

An ER random graph instance
with a vertex size $N$ and a mean degree $c$
can be constructed by adding $M (\equiv cN / 2)$ edges to an empty graph with only $N$ vertices
by connecting pairs of randomly chosen distinct vertices.

On ER random graphs,
we have the degree distributions as
\begin{eqnarray}
P(k)
&&
= e ^{- c} \frac{c ^{k}}{k!}, \\
Q(k)
&&
= e^{- c} \frac{c^{k - 1}}{(k - 1)!}.
\end{eqnarray}
We have simplified equations as
\begin{eqnarray}
\label{eq:alpha_er}
\alpha
& &
= e^{- c (1 - \beta)}, \\
\label{eq:beta_er}
\beta
& &
= 1 - e ^{- c \alpha}, \\
\label{eq:n_er}
n
& &
= e ^{- c \alpha}
- e^{- c (1 - \beta)}
- c \alpha (1 - \alpha - \beta), \\
\label{eq:w_er}
w
& &
= 1 - e ^{- c \alpha} - \frac {1}{2} c \alpha ^{2}.
\end{eqnarray}

\subsection{Regular random graphs}

For dRR graphs after an edge dilution scheme with a parameter $\rho$
on RR graphs with an original degree $K$,
we have degree distributions
\begin{eqnarray}
P(k)
&&
= \left(\begin{array}{c} K \\ k \end{array}\right)
\rho ^{k} (1 - \rho)^{K - k}, \\
Q(k)
&&
= \left(\begin{array}{c} K - 1 \\ k - 1 \end{array}\right)
\rho ^{k - 1} (1 - \rho)^{K - k}.
\end{eqnarray}
%
We further have simplified equations as
\begin{eqnarray}
\alpha
&&
= (1 - \rho (1 - \beta))^{K - 1}, \\
\beta
&&
= 1 - (1 - \rho \alpha)^{K - 1}, \\
n
&&
= (1 - \rho \alpha)^{K}
- (1 - \rho (1 - \beta))^{K}
- \rho K \alpha (1 - \alpha - \beta), \\
w
&&
= 1 - (1 - \rho \alpha )^{K}  - \frac {1}{2} \rho K \alpha ^{2}.
\end{eqnarray}

\subsection{Scale-free networks with the configurational model}

First we show how to construct scale-free (SF) networks with the configurational model.
The key graph parameters here are:
$N$ the vertex size,
 $\gamma$ the degree exponent,
$k_{\rm{min}}$ the minimal degree, and
$k_{\rm{max}}$ the maximal degree.
For simplicity,
 in this paper we set $k_{\rm{min}} \ge 2$ and $k_{\rm{max}} = \sqrt N$.
The graph generation is as follows.
(1) A sequence of degrees is generated
with $N(k) = N P_{0}(k)$ with $k_{\rm{min}} \le k \le k_{\rm{max}}$ and
$P_{0}(k) = k^{- \gamma} / \sum _{m = k_{\rm{min}}}^{k_{\rm{max}}} m^{- \gamma}$.
Thus we have a size of free studs (half-edges) $E = \sum _{k = k_{\rm{min}}}^{k_{\rm{max}}} k N(k)$.
(2) For any vertex $i$,
we assign its degree $k_{i}$ sampled from the distribution $N(k)$.
Thus we have an empty graph configuration with $N$ vertices while any vertex $i$ has $k_{i}$ free studs.
If $E$ is odd,
we can randomly chose a vertex $i$ and increase $k_{i}$ by $1$
to make $E$ even.
Thus we have the mean degree of the graph instance as $c_{0} = E / N$.
(3) To add edges into an empty graph,
we randomly choose two free studs adjacent to two distinct vertices to establish a proper edge.
In the case of any self-loop (a vertex connected to itself)
or multi-edge (two vertices sharing two edges),
we can randomly choose an established edge and exchange their studs
to reestablish two new edges.
The edge construction procedure is repeated until there is no free stud.

On a SF network with $k_{\rm{min}} \ge 2$,
an edge dilution scheme with a parameter $\rho$ can be applied
to trigger the GLR procedure.
For the diluted SF networks,
we have the degree distributions as
\begin{eqnarray}
P(k)
&&
= \frac{1}{\sum _{m} m^{- \gamma}}
\sum _{t}
t^{- \gamma}
 \left(\begin{array}{c} t \\ k \end{array}\right)
\rho ^{k}
(1 - \rho)^{t - k}, \\
Q(k)
&&
= \frac{1}{\sum _{m} m^{1 - \gamma}}
\sum _{t}
t^{1 - \gamma}
 \left(\begin{array}{c} t - 1 \\ k - 1 \end{array}\right)
\rho ^{k - 1}
(1 - \rho)^{t - k},
\end{eqnarray}
in which
$\sum _{m}$is short-handed for $\sum _{m = k_{\rm{min}}} ^{k_{\rm{max}}}$ and
$\sum _{t}$ for $\sum _{t = \max \{k, k_{\rm{min}}\}}^{k_{\rm{max}}}$.
We thus have the simplified equations as
\begin{eqnarray}
\alpha
&&
= \frac {1}{\sum _{m} m ^{1 - \gamma}}
\sum _{k = 1}^{\infty}
\sum _{t}
t^{1 - \gamma}
\left(\begin{array}{c} t - 1 \\ k - 1 \end{array}\right)
\rho ^{k - 1}
(1 - \rho)^{t - k}
\beta^{k - 1}, \\
 \beta
&&
= 1 - \frac {1}{\sum _{m} m ^{1 - \gamma}}
\sum _{k = 1}^{\infty}
\sum _{t}
t^{1 - \gamma}
\left(\begin{array}{c} t - 1 \\ k - 1 \end{array}\right)
\rho ^{k - 1}
(1 - \rho)^{t - k}
(1 - \alpha)^{k - 1}, \\
n
&&
= \frac {1}{\sum _{m} m ^{- \gamma}}
\sum _{k = 0}^{\infty}
\sum _{t}
t^{- \gamma}
\left(\begin{array}{c} t \\ k \end{array}\right)
\rho ^{k}
(1 - \rho)^{t - k}
[(1 - \alpha)^{k} - \beta^{k}]
- \rho c_{0} \alpha (1 - \alpha - \beta), \\
w
&&
= 1 - \frac {1}{\sum _{m} m ^{- \gamma}}
\sum _{k = 0}^{\infty}
\sum _{t}
t^{- \gamma}
\left(\begin{array}{c} t \\ k \end{array}\right)
\rho ^{k}
(1 - \rho)^{t - k} (1 - \alpha)^{k}
- \frac{1}{2} \rho c_{0} \alpha^2.
\end{eqnarray}

\subsection{Scale-free networks with the static model}

First we show how to construct a SF network instance with the static model.
We denote
$\gamma$ as the degree exponent, $c$ the mean degree, and $N$ the vertex size.
We define an auxiliary coefficient $\xi \equiv 1 / (\gamma - 1)$.
In the graph construction,
we follow such procedures:
(1) on an empty graph with only $N$ vertices,
each vertex $i$ is assigned with a weight $w_{i} =  i ^{- \xi}$,
where $i \in \{1, ..., N\}$ is the index for each vertex;
(2) in a single step of edge generation,
two distinct vertices are chosen with probabilities proportional to their respective weights
and are further connected;
in such a way,
a number of $M (\equiv c N / 2)$ edges
are added into the empty graph.

For the SF networks generated with the static model,
we have the degree distributions
\begin{eqnarray}
\label{eq:Pk_sm}
P(k)
&&
= \frac {1}{\xi} \frac {(c (1 - \xi))^{k}}{k!}
E _{- k + 1 + 1 / \xi} (c(1 - \xi)), \\
Q(k)
&&
= \frac {1 - \xi}{\xi} \frac {(c (1 - \xi))^{k - 1}}{(k - 1)!}
E_{- k + 1 + 1 / \xi} (c (1 - \xi)),
\end{eqnarray}
while $E _{a}(x) \equiv \int _{1}^{\infty} {\rm d}t e^{- x t} t ^{- a}$.
In the case of large $k$,
we asymptotically have $P(k) \propto k ^{- \gamma}$.
To solve the special functions,
we reformulate the general exponential integral function as
$E _{a}(x) \equiv x^{a - 1} \Gamma (1 - a, x)$,
in which $\Gamma (\cdot , x)$ with $x > 0$ is an upper incomplete gamma function.
We use the GNU Scientific Library
\cite{GSL}
to calculate $\Gamma (1 - a, x)$, thus $E_{a}(x)$.
We have simplified equations as
\begin{eqnarray}
\alpha
&&
= \frac {1 - \xi}{\xi} E_{\frac {1}{\xi}} (c (1 - \xi) (1 - \beta)), \\
\beta
&&
= 1 - \frac {1 - \xi}{\xi} E_{\frac {1}{\xi}} (c (1 - \xi) \alpha), \\
n
&&
= \frac {1}{\xi} E_{1 + \frac {1}{\xi}} (c (1 - \xi) \alpha)
- \frac {1}{\xi} E_{1 + \frac {1}{\xi}} (c (1 - \xi) (1 - \beta))
- c \alpha (1 - \alpha - \beta), \\
w
&&
= 1 - \frac {1}{\xi} E_{1 + \frac {1}{\xi}} (c (1 - \xi) \alpha)
- \frac {1}{2} c \alpha ^{2}.
\end{eqnarray}

\section{Appendix D: Description of the real-world network dataset}

In the dataset,
there are $37$ network instances in $12$ categories.
Most of the large networks in the datase are from SNAP Datasets
\cite{SNAP}.
In Table (\ref{tab:real_1}) - (\ref{tab:real_3}),
for each network instance,
we show its category and name,
a brief description,
its edge type,
its vertex size $N$, and
its size of undirected edge or directed arc $M$.

For any directed network instance,
we only retain its connection pattern to derive its undirected counterpart.
As a preprocessing on the dataset,
we remove self-loops (vertices connecting to themselves)
and merge multi-edges (pairs of vertices connected by multiple edges)
in each network instance.

\begin{table}
\caption{1/3 of the dataset.}
\label{tab:real_1}
\begin{center}
\begin{tabular*}{\linewidth}{p{3.5cm}p{7cm}p{2cm}p{1.5cm}p{1.5cm}}
\hline
  Name 
  & Description
  & Edge type
  & $N$
  & $M$ \\
 \hline
 Regulatory&&&& \\
 EGFR \cite{Fiedler.Mochizuki.Kurosawa.Saito-JDynDiffEquat-2013a}
 & Simplified signal transduction network of EGF receptors.
 & directed
 & $61$
 & $112$ \\
 {\em E. coli}  \cite{Mangan.Alon-PNAS-2003}
 &  Transcriptional regulatory network of {\em E. coli}.
 & directed
 & $418$
 & $519$ \\
 {\em S. cerevisiae} \cite{Alon.etal-Science-2002}
 & Transcriptional regulatory network of {\em S. cerevisiae}.
 & directed
 & $688$
 & $1,079$ \\
 \hline
 PPI &&&&\\
 PPI \cite{Vinayagam.etal-ScienceSignaling-2011}
 & Protein-protein interaction network.
 & directed
 & $6,339$
 & $34,814$ \\
 \hline
 Metabolic &&&&\\
 {\em C. elegans} \cite{Jeong.etal-Nature-2000}
 & Metabolic network of {\em C. elegans}.
 & directed
 & $1,469$
 & $3,447$ \\
 {\em S. cerevisiae} \cite{Jeong.etal-Nature-2000}
 & Metabolic network of {\em S. cerevisiae}.
 & directed
 & $1,511$
 & $3,833$ \\
 {\em E. coli} \cite{Jeong.etal-Nature-2000}
 & Metabolic network of {\em E. coli}.
 & directed
 & $2,275$
 & $5,763$ \\
 \hline
 Neuronal &&&&\\
 {\em C. elegans} \cite{Watts.Strogatz-Nature-1998}
 & Neural network of {\em C. elegans}.
 & directed
 & $297$
 & $2,345$ \\
 \hline
 Food webs &&&&\\
 Maspalomas \cite{foodweb-Maspalomas-1999}
 & Food web in Charca de Maspalomas.
 & directed
 & $24$
 & $82$ \\
 Chesapeake \cite{foodweb-Chesapeake-1989}
 & Food web in Chesapeake Bay.
 & directed
 & $39$
 & $176$ \\
 St Marks \cite{foodweb-StMarks-1998}
 & Food web in St. Marks River Estuary.
 & directed
 & $54$
 & $353$ \\
 Everglades \cite{foodweb-Everglades-2000}
 & Food web in  Everglades Graminoid Marshes. 
 & directed
 & $69$
 & $911$ \\
 Florida Bay \cite{foodweb-Florida-1998}
 & Food web in Florida Bay. 
 & directed
 & $128$
 & $2,106$ \\
 \hline
 Electronic &&&&\\
 s208 \cite{Alon.etal-Science-2002}
 & Electronic sequential logic circuits.
 & directed
 & $122$
 & $189$ \\
 s420 \cite{Alon.etal-Science-2002}
 & Same as above.
 & directed
 & $252$
 & $399$ \\
 s838 \cite{Alon.etal-Science-2002}
 & Same as above.
 & directed
 & $512$
 & $819$ \\
 \hline
 Ownership &&&&\\
 USCorp \cite{Norlen.etal-PITS14-2002}
 & Ownership network of US corporations.
 & directed
 & $7,253$
 & $6,724$ \\
\hline
\end{tabular*}
\end{center}
\end{table}

\begin{table}
\caption{2/3 of the dataset.}
\label{tab:real_2}
\begin{center}
\begin{tabular*}{\linewidth}{p{3.5cm}p{7cm}p{2cm}p{1.5cm}p{1.5cm}}
\hline
  Name 
  & Description
  & Edge type
  & $N$
  & $M$ \\
 \hline
Collaboration &&&&\\
 GrQc \cite{Leskovec.Kleinberg.Faloutsos-TransKDD-2007}
 & Collaboration network of Arxiv General Relativity.
 & undirected
 & $5,241$
 & $14,484$ \\
 HepTh \cite{Leskovec.Kleinberg.Faloutsos-TransKDD-2007}
 & Collaboration network of Arxiv High Energy Physics Theory.
 & undirected
 & $9,875$
 & $25,973$ \\
 HepPh \cite{Leskovec.Kleinberg.Faloutsos-TransKDD-2007}
 & Collaboration network of Arxiv High Energy Physics.
 & undirected
 & $12,006$
 & $118,489$ \\
 AstroPh \cite{Leskovec.Kleinberg.Faloutsos-TransKDD-2007}
 & Collaboration network of Arxiv Astro Physics.
 & undirected
 & $18,771$
 & $198,050$ \\
 CondMat \cite{Leskovec.Kleinberg.Faloutsos-TransKDD-2007}
 & Collaboration network of Arxiv Condensed Matter.
 & undirected
 & $23,133$
 & $93,439$ \\
\hline
 Citation &&&& \\
 HepTh \cite{Leskovec.Kleinberg.Faloutsos-SIGKDD-2005}
 & Citation network in HEP-TH category of ArXiv.
 & directed
 & $27,769$
 & $352,768$ \\
 HepPh \cite{Leskovec.Kleinberg.Faloutsos-SIGKDD-2005}
 & Citation network in HEP-PH category of ArXiv.
 & directed
 & $34,546$
 & $421,534$ \\
\hline
 Internet P2P &&&&\\
 Gnutella04 \cite{Ripeanu.Iamnitchi.Foster-IEEEInternetComputing-2002, Leskovec.Kleinberg.Faloutsos-TransKDD-2007}
 & Gnutella peer-to-peer network from August 4, 2002. 
 & directed
 & $10,876$
 & $39,994$ \\
 Gnutella30 \cite{Ripeanu.Iamnitchi.Foster-IEEEInternetComputing-2002, Leskovec.Kleinberg.Faloutsos-TransKDD-2007}
 & Gnutella peer-to-peer network from August 30, 2002.
 & directed
 & $36,682$
 & $88,328$ \\
 Gnutella31 \cite{Ripeanu.Iamnitchi.Foster-IEEEInternetComputing-2002, Leskovec.Kleinberg.Faloutsos-TransKDD-2007}
 & Gnutella peer-to-peer network from August 31, 2002.
 & directed
 & $62,586$
 & $147,892$ \\
\hline
 Web graphs &&&&\\
 NotreDame \cite{Albert.Jeong.Barabasi-Nature-1999}
 & Web graph of Notre Dame.
 & directed
 & $325,729$
 & $1,469,679$ \\
 Stanford \cite{Leskovec.etal-InternetMath-2009}
 & Web graph of Stanford.edu. 
 & directed
 & $281,903$
 & $2,312,497$ \\
 Google \cite{Leskovec.etal-InternetMath-2009}
 & Web graph from Google.
 & directed
 & $875,713$
 & $5,105,039$ \\
\hline
\end{tabular*}
\end{center}
\end{table}

\begin{table}
\caption{3/3 of the dataset.}
\label{tab:real_3}
\begin{center}
\begin{tabular*}{\linewidth}{p{3.5cm}p{7cm}p{2cm}p{1.5cm}p{1.5cm}}
\hline
  Name 
  & Description
  & Edge type
  & $N$
  & $M$ \\
 \hline
 Social &&&&\\
 Karate \cite{Zachary-JourAnthropResearch-1977}
 & Social network of friendship between members in a Karate club.
 & undirected
 & $34$
 & $78$ \\
 Dolphins \cite{Lusseus.etal-BehavEcoSocio-2003}
 & Social network of frequent associations in a dolphin community.
 & undirected
 & $62$
 & $159$ \\
 Football \cite{Girvan.Newman-PNAS-2002}
 & Network of American football games.
 & undirected
 & $115$
 & $613$ \\
 Enron \cite{Klimt.Yang-2004, Leskovec.etal-InternetMath-2009}
 & Email communication network from Enron.
 & undirected
 & $36,692$
 & $183,831$ \\
 WikiVote \cite{Leskovec.Huttenlocher.Kleinberg-SIGCHI-2010, Leskovec.Huttenlocher.Kleinberg-ICWWW-2010}
 & Who-vote-whom network of Wikipedia users. 
 & directed
 & $7,115$
 & $103,689$ \\
 Epinions \cite{Richardson.etal-ISWC-2003}
 & Who-trust-whom network of Epinions.com users. 
 & directed
 & $75,879$
 & $508,837$ \\
 EuAll \cite{Leskovec.Kleinberg.Faloutsos-TransKDD-2007}
 & Email network from a EU research institution. 
 & directed
 & $265,009$
 & $418,956$ \\
\hline
\end{tabular*}
\end{center}
\end{table}

\end{document}